\documentclass[12pt]{article}

\usepackage{amssymb}
\usepackage{hyperref}
\usepackage{amsmath}
\usepackage{graphicx}
\usepackage{subfig}

\usepackage{color}

\newcommand {\apgt} {\ {\raise-.5ex\hbox{$\buildrel>\over\sim$}}\ }

\begin{document}

\begin{titlepage}

\vspace{-4cm}

\title{
   {\LARGE $B$-$L$ Cosmic Strings in Heterotic \\[.1in]
    Standard Models\\[.5cm]  }}
                       
\author{{\bf
   Tamaz Brelidze$^{1}$ 
   and Burt A.~Ovrut$^{1,2}$}\\[5mm]
   {\it $^1$Department of Physics, University of Pennsylvania} \\
   {\it Philadelphia, PA 19104--6396}\\[2mm]
   {\it $^2$School of Natural Sciences, The Institute for Advanced Study} \\
   {\it Princeton, New Jersey, 08540}} 
\date{}

\maketitle

\begin{abstract}
\noindent $E_{8} \times E_{8}$ heterotic string and M-theory, when compactified on smooth Calabi-Yau manifolds with $SU(4)$ vector bundles, can give rise to softly broken $N=1$ supersymmetric theories with the exact matter spectrum of the MSSM, including three right-handed neutrinos
and one Higgs-Higgs conjugate pair of supermultiplets. These 
vacua have the $SU(3)_{C} \times SU(2)_{L} \times U(1)_{Y}$ gauge group of the standard model augmented by an additional gauged $U(1)_{B-L}$. Their minimal content requires that the $B$-$L$ symmetry be spontaneously broken by a vacuum expectation value of at least one right-handed sneutrino. The soft supersymmetry breaking operators can induce radiative breaking of the $B$-$L$ gauge symmetry with an acceptable $B$-$L$/electroweak hierarchy. In this paper, it is shown that $U(1)_{B-L}$ cosmic strings  occur in this context, potentially with both bosonic and fermionic superconductivity. 
We present a numerical analysis that demonstrates that boson condensates can, in principle, form for theories of this type. However, the weak Yukawa and gauge couplings of the right-handed sneutrino suggests that bosonic superconductivity will not occur in the simplest vacua in this context. The electroweak phase transition also disallows fermion superconductivity, although substantial bound state fermion currents can exist.

\vspace{.3in}
\noindent
\end{abstract}

\thispagestyle{empty}

\end{titlepage}

\section{Introduction}

Heterotic string and $M$-theory when compactified on smooth geometric and vector bundle backgrounds~\cite{a}-\cite{h} can give rise to ``heterotic standard models''~\cite{exact,donagi,lukas}; that is, four-dimensional $N=1$ supersymmetric theories with exactly the matter and Higgs spectrum of the MSSM. Supersymmetry can be spontaneously broken by non-perturbative effects in the hidden sector.
Integrating out this sector, the low-energy theory contains ``soft'' supersymmetry breaking operators whose generic form
is well-known~\cite{Xaa}-\cite{Xe}. To be phenomenologically viable, any such theory must have two properties: 1) three right-handed neutrino chiral multiplets, one per family, and 2) ``matter parity'', a discrete ${\bf {Z}}_{2}$ symmetry which prohibits too rapid baryon and lepton number violating processes~\cite{Ua}-\cite{Ud}. 

These two properties are most easily satified in heterotic standard models constructed using vector bundles with $SU(4)$ structure group~\cite{braun1}-\cite{donagi2}. In addition to the MSSM spectrum, such vacua  have three right-handed neutrino chiral multiplets, thus
satisfying the first property. The low-energy gauge group also contains the $SU(3)_{C} \times SU(2)_{L} \times U(1)_{Y}$ gauge group of the standard model augmented, however, by a gauged $U(1)_{B-L}$ factor. This contains matter parity as a discrete subgroup. If the $B$-$L$ symmetry could be spontaneously broken to its matter parity subgroup, then the second property would be satisfied as well. However, this is not possible in smooth heterotic compactifications since the necessary $3(B-L)$-even multiplets are disallowed as zero-modes. A second solution is to have $U(1)_{B-L}$ radiatively broken at low energy, not too far above the electroweak scale. It would then act as a custodial symmetry for matter parity, suppressing baryon and lepton violating decays yet not unduly affecting electroweak physics.

In several recent papers~\cite{mike1,mike2}, it was shown using a quasi-analytic solution to the renormalization group equations (RGEs) that this can indeed occur for a range of initial soft breaking parameters. Scaling down from the compactifcation mass, the gauged $U(1)_{B-L}$ is first spontaneously broken by a non-zero vacuum expectation value (VEV) of the third family right-handed sneutrino. This is followed by radiative VEVs developing in the Higgs fields which induce an electroweak phase transition. The $B$-$L$/electroweak hierarchy was found to be of ${\cal{O}}(10)$-${\cal{O}}(100)$. Recently, these results have been expanded to a much wider range of initial soft parameters using a completely numerical solution of the RGEs. This work will appear elsewhere~\cite{mike3}. Here, we simply note that this expanded range of parameters leads to three distinct possibilities for the soft mass squared parameters of squarks/sleptons at the electroweak scale. In addition to the negative third family right-handed sneutrino mass,  1) all such parameters are positive, 2) all are positive with the exception of a right-handed charged slepton and 3) all are positive with the exception of a left-handed squark. Each possibility can play an interesting role in cosmology. 

The starting point of this paper is the assumption that smooth heterotic compactifications with $SU(4)$ structure group are potentially phenomenologically viable theories for low-energy particle physics. The distinct signature of this type of vacuum is that a gauged $U(1)_{B-L}$ symmetry is spontaneously broken at a low scale, not too far above the electroweak phase transition. As is well-known~\cite
{Abrikos, Olsen}, the breaking of a gauged Abelian symetry can lead to topologically stable cosmic strings. In principle, these strings can exhibit a wide variety of observable cosmological phenomena~\cite{Chudnovsky}-\cite{BattWesley}. However, much of the analysis of cosmic strings has been carried out within the context of  grand unified theories or specially constructed supersymmetric models whose spectra contain fields in addition to those of the MSSM with right-handed neutrinos~\cite{Everett}-\cite{Trodden}. Furthermore, the coupling parameters associated with these fields are not constrained and can be assumed to be sufficiently large. As a rule, it is these extra fields that induce the potentially observable phenomena, such as bosonic or fermionic superconductivity~\cite{MarkBrand}. In smooth $B$-$L$ MSSM heterotic compactifications, there are no additional fields. The breaking of $U(1)_{B-L}$ and electroweak symmetry is accomplished via radiative expectation values for a right-handed sneutrino and Higgs fields respectively. As a consequence, the relevant parameters in this theory are those of the MSSM and, hence, tightly constrained by phenomenology. It follows that the existence and properties of cosmic strings in this context are severely restricted. 

In this paper, we analyze 
cosmic strings in the $B$-$L$ MSSM theory. We show that such strings can indeed exist but are restricted to be BPS solutions at the critical  boundary between Type I and Type II superconductors. There is a stable minimum of the scalar potential in which 1) $B$-$L$ is broken by a VEV $\langle\nu_{3} \rangle$ of the third family right-handed sneutrino and 
2) electroweak symmetry is broken by Higgs expectation values  $\langle H^{0} \rangle$,
$\langle {\bar{H}}^{0} \rangle$. At this minimum, all other scalar fields have positive squared masses. Some, however, specifically right-handed charged sleptons and left-chiral squarks, have effective masses $\langle m^{2} \rangle =m^{2}+c\langle \nu_{3} \rangle^{2}$ with positive coefficient $c$, where $m^{2}$ is the associated soft breaking mass parameter. Three possibilites then arise. First, all $m^{2}$ parameters might be positive, making the potential energy at the origin of field space a minimum in all but the third family sneutrino direction. Second, a charged right-handed slepton could have a negative $m^{2}$ parameter, in addition to the right-handed sneutrino. Although the $B$-$L$/electroweak vacuum remains the minimum, this destabilizes the slepton direction in the core of the cosmic string, potentially leading to a charge breaking condensate and bosonic superconductivity~\cite{Witten}. The third possibility is that the soft mass squared parameter for a squark becomes negative, potentially leading to a charge and color breaking superconducting condensate. We have shown in~\cite{mike3} that each of these types of vacua is possible for a given range of initial soft parameters. 
From both the cosmological and phenomenological point of view, it is of interest to see if bosonic supercondictivity can occur in $B$-$L$ MSSM cosmic strings. To explore this, 
we study a generic class of theories that arise in this context. Using a numerical analysis, a bound is derived that must be satisfied to allow the formation of a non-zero condensate and, hence, bosonic superconductivity. This analysis is then applied to  the most straightforward $B$-$L$ MSSM vacua and cosmic strings using simplifying assumptions. We find that the right-handed sneutrino Yukawa parameter and the $g_{B-L}$ gauge coupling are too small to permit this essential constraint to be satisfied. We conclude that at least the simplest $B$-$L$ MSSM cosmic strings do not exhibit bosonic superconductivity.

$B$-$L$ MSSM cosmic strings may also exhibit superconductivity induced by fermionic zero-modes in the string core~\cite{Kiskis, Jackiw, Witten}. The fact that the gauged $U(1)_{B-L}$ extension of the MSSM is rendered anomaly free by the inclusion of three families of right-handed neutrino chiral multiplets plays an important role here. The cosmic string initially develops as a non-zero $n$-fold winding of $\langle \nu_{3} \rangle$ around some line in space. This couples directly to the left-chiral tauon $\psi_{E^{-}}$ and the chargino $\psi_{H^{+}}$, forming left-moving fermion zero-modes on the string worldsheet. However, anomaly cancellation requires the appearance of right-moving fermionic modes, whose identity in the $B$-$L$ MSSM context  is not self-evident. We show that the third family left-handed sneutrino develops a small VEV $\langle N_{3} \rangle$ following the electroweak phase transition. This wraps the core of the cosmic string with winding $-n$, opposite that that of $\langle \nu_{3} \rangle$. The right-chiral tauon $\psi_{e^{+}}$ and chargino $\psi_{H^{-}}$ couple to this field, inducing right-handed zero-modes which cancel all worldsheet anomalies. Thus, there is potential fermionic superconductivity in the cosmic string. We conclude, however, that the electroweak phase transition will, in general, render these fermionic currents unobservable~\cite{HillWidow}-\cite{PhaseTr}.

The paper is structured as follows. In Section 2, we briefly review the spectrum, superpotential and potential energy of the softly broken $B$-$L$ MSSM theory. The structure of the $B$-$L$ and electroweak breaking vacuum is then presented in Section 3, including the effective scalar masses at this minimum and the $B$-$L$/electroweak hierarchy. Section 4 is devoted to showing that the winding of the $B$-$L$ charged right-handed sneutrino VEV around the origin leads to a cosmic string with critical coupling. The allowed patterns of soft scalar masses at the core of the cosmic string are discussed in Section 5. For each case, the stability criterion for a scalar condensate to develop in the string core, and, hence, for the string to be potentially superconducting, is derived. These criteria are then analyzed using a numerical analysis presented in the Appendix. Finally, in Section 6 
we discuss potential fermionic zero-modes, show how the anomaly freedom of the $B$-$L$ MSSM theory leads to appropriate left- and right-moving modes and present the constraints imposed on these currents by the breaking of $B$-$L$ via the right-handed sneutrino.

\section{The $N=1$ Supersymmetric Theory}

We consider an $N=1$ supersymmetric theory with gauge group
\begin{equation}
G=SU(3)_{C} \times SU(2)_{L} \times U(1)_{Y} \times U(1)_{B-L}
\label{1}
\end{equation}
and the associated vector superfields. The gauge parameters are denoted $g_{3}$, $g_{2}$, $g_{Y}$ and $g_{B-L}$ respectively. The matter spectrum consists of three families of quark/lepton chiral superfields, each family with a {\it right-handed neutrino}. They transform under the gauge group in the standard manner as
\begin{equation}
Q_{i}=({\bf 3},{\bf 2},1/3,1/3), \quad u_{i}=({\bf \bar{3}}, {\bf 1}, -4/3, -1/3), \quad d_{i}=({\bf \bar{3}}, {\bf 1}, 2/3, -1/3)
\label{2}
\end{equation}
for the left and right-handed quarks and
\begin{equation}
L_{i}=({\bf 1},{\bf 2},-1,-1), \quad \nu_{i}=({\bf 1}, {\bf 1}, 0, 1), \quad e_{i}=({\bf 1}, {\bf 1}, 2, 1)
\label{3}
\end{equation}
for the left  and right-handed leptons, where $i=1,2,3$. In addition, the spectrum has one pair of Higgs-Higgs conjugate chiral superfields transforming as
\begin{equation}
H=({\bf 1},{\bf 2},1,0), \qquad \bar{H}=({\bf 1},{\bf 2}, -1,0).
\label{4}
\end{equation}
When necessary, the left-handed $SU(2)_{L}$ doublets will be written as 
\begin{equation}
Q_{i}=(U_{i}, D_{i}), \quad L_{i}=(N_{i}, E_{i}), \quad H=(H^{+},H^{0}), \quad \bar{H}=({\bar{H}}^{0}, {\bar{H}}^{-}).
\label{5}
\end{equation}
There are {\it no other fields in the spectrum}. The three right-handed neutrino chiral multiplets render this $U(1)_{B-L}$ extension of the  MSSM anomaly free.

The supersymmetric potential energy is given by the sum over the modulus squared of the $F$
and $D$-terms. The $F$-terms are determined from the superpotential
\begin{equation}
W=\mu H\bar{H} +{\sum_{i=1}^{3}}\left(\lambda_{u, i} Q_{i}Hu_{i}+\lambda_{d, i} Q_{i}\bar{H}d_{i}+\lambda_{\nu, i} L_{i}H\nu_{i}+\lambda_{e, i} L_{i}\bar{H}e_{i}\right) ,
\label{6}
\end{equation}
where we assume a mass-diagonal basis for simplicity.
An innocuous mixing term of the form $L_{i}H$
as well as the dangerous lepton and baryon number violating interactions
\begin{equation}
L_{i}L_{j}e_{k}, \quad L_{i}Q_{j}d_{k}, \quad u_{i}d_{j}d_{k}
\label{7}
\end{equation}
are disallowed by the $U(1)_{B-L}$ gauge symmetry. 
The $SU(3)_{C}$ and $SU(2)_{L}$ $D$-terms are of standard form. The $U(1)_{Y}$ 
and $U(1)_{B-L}$ $D$-terms are 
\begin{equation}
D_{Y}= g_{Y}{\phi}_{A}^{\dagger}\left({\bf{Y}\rm}/2\right)_{AB}{\phi}_{B}
\label{9}
\end{equation}
and 
\begin{equation}
D_{B-L}= g_{B-L}{\phi}_{A}^{\dagger}\left({\bf{Y_{B-L}}\rm}\right)_{AB}{\phi}_{B}
\label{10}
\end{equation}
respectively, where index $A$ runs over all scalar fields ${\phi}_{A}$. 
In the $D$-eliminated formalism, any Fayet-Iliopoulos parameters can be consistently absorbed into the definition of the soft supersymmetry breaking scalar masses. Hence, they do not appear in \eqref{9} and \eqref{10}.

In addition to supersymmetric interactions, the potential energy also contains explicit soft supersymmetry violating terms. 
This breaking can arise in either $F$-terms, $D$-terms or both in the hidden sector. 
We will restrict our discussion to soft supersymmetry breaking scalar interactions arising exclusively from  $F$-terms. Their form is well-known and, in the present context, given by \cite{Xaa}-\cite{Xe}
\begin{equation}
V_{\rm soft}=V_{2s}+V_{3s},
\label{11}
\end{equation}
where $V_{2s}$ are scalar mass terms 
\begin{eqnarray}
&& V_{2s}  = { \sum_{i=1}^{3}} (m^{2}_{Q_{i}}|{Q}_{i}|^{2}+m^{2}_{u_{i}}|{u}_{i}|^{2}+
                        m^{2}_{d_{i}}|{d}_{i}|^{2}  +m^{2}_{L_{i}}|{L}_{i}|^{2}  +m^{2}_{\nu_{i}}|{\nu}_{i}|^{2} \nonumber    \\ 
&& \qquad \qquad   +m^{2}_{e_{i}}|{e}_{i}|^{2} +m_{H}^{2}|H|^{2} +m_{\bar{H}}^{2}|\bar{H}|^{2})-(BH\bar{H}+hc) \label{12}  
\end{eqnarray}
and $V_{3s}$ are the scalar cubic couplings
\begin{equation}
 V_{3s}=\sum_{i=1}^{3} (A_{u_{i}} {Q}_{i}H{u}_{i} +A_{d_{i}} {Q}_{i}{\bar{H}}{b}_{i} +A_{\nu_{i}} {L}_{i}H{\tilde{\nu}}_{i}+A_{e_{i}} {L}_{i}{\bar{H}}{e}_{i} +{\rm hc}) .
\label{13}
\end{equation}
We choose the parameters in (\ref{12}) and (\ref{13}) to be flavor-diagonal. 

\section{The $B$-$L$/Electroweak Hierarchy}

In~\cite{mike1,mike2} a detailed one-loop renormalization group analysis of this theory was carried out. In that analysis, $tan\beta$ was limited to $6.32 \leq tan\beta \leq 40$ and a specific range of initial parameters  near the gauge unification scale $M_{u} \simeq 3  \times 10^{16} GeV$ was chosen so as to allow a quasi-analytic solution of the RGEs. Here, we simply present the results.  Subject to realistic, but constrainted, assumptions about the soft breaking parameters, it was shown that a hierarchy of radiative symmetry breaking takes place. 

First, at an energy scale of $\sim TeV$ the third family right-handed sneutrino soft mass parameter is negative; that is, $m_{\nu_{3}}^{2}<0$. It follows that this sneutrino acquires a vacuum expectation value (VEV) 
\begin{equation}
\langle \nu_{3} \rangle^{2}= -\frac{m_{\nu_{3}}^{2}}{g_{B-L}^{2}} \ .
\label{snow1}
\end{equation}
Furthermore, evaluated at $\langle \nu_{3} \rangle$ all other scalars, including the Higgs fields, have vanishing VEVs. Therefore, this is a stable vacuum which spontaneously breaks $U(1)_{B-L}$ while preserving the remaining 
$SU(3)_{C} \times SU(2)_{L} \times U(1)_{Y}$ gauge symmetry. The Higgs effect then leads to identical masses for the $B$-$L$ vector boson and the radial real scalar $\delta \nu_3$ given by 
\begin{equation}
M_{A_{B-L}}=m_{\delta \nu_3}=\sqrt{2}g_{B-L}\langle \nu_{3} \rangle \ .
\label{snow2}
\end{equation}
It is of interest to recall the expressions for the slepton/squark masses at this minimum of the potential. They were found to be
\begin{eqnarray}
&& \quad  \langle m_{L_{i}}^{2} \rangle  =  m_{L_{i}\rm}^{2}-g_{B-L}^{2}\langle \nu_{3} \rangle^{2}
\nonumber \\
\langle m_{\nu_{1,2}}^{2} \rangle &=& m_{\nu_{1,2}\rm}^{2}+g_{B-L}^{2}\langle \nu_{3} \rangle^{2}, \quad \langle m_{e_{i}}^{2} \rangle = m_{e_{i}\rm}^{2}+g_{B-L}^{2} \langle \nu_{3} \rangle^{2} 
\label{snow3} 
\end{eqnarray}
and
\begin{eqnarray}
&& \quad  \langle m_{Q_{i}}^{2} \rangle = m_{Q_{i}}^{2}+\frac{1}{3} g_{B-L}^{2} \langle \nu_{3} \rangle^{2}
\nonumber \\
\langle m_{u_{i}}^{2} \rangle &= & m_{u_{i}}^{2}-\frac{1}{3} g_{B-L}^{2} \langle \nu_{3} \rangle^{2}, \quad  \langle m_{d_{i}}^{2} \rangle = m_{d_{i}}^{2}-\frac{1}{3} g_{B-L}^{2} \langle \nu_{3} \rangle^{2} 
\label{snow4} 
\end{eqnarray}
for $i=1,2,3$.
Note from the minus sign in the expressions for $\langle m_{L_{i}}^{2} \rangle$ and 
$\langle m_{u_{i}}^{2} \rangle$, $\langle m_{d_{i}}^{2} \rangle$ that for this to be a stable vacuum the soft mass parameters $m_{L_{i}}^{2}$ and $m_{u_{i}}^{2}$, $m_{d_{i}}^{2}$ must always be positive at the $B$-$L$ scale. This was shown to be the case. However, the same is not required for the 
$m_{\nu_{1,2}}^{2}$, $m_{e_{i}}^{2}$ and $m_{Q_{i}}^{2}$ parameters. These can become negative at the $B$-$L$ scale as long as $\langle m_{\nu_{1,2}}^{2} \rangle$, $\langle m_{e_{i}}^{2} \rangle$ and $ \langle m_{Q_{i}}^{2} \rangle$ are positive. This has important implications for bosonic superconductivity, as we will discuss in Section 5. For simplicity, we will always take 
$m_{\nu_{1,2}}^{2}$ to be positive at any scale, as was done in~\cite{mike1,mike2}.

Second, scale all parameters down to $\sim 10^{2}GeV$. Here, one of the diagonalized Higgs soft masses, indicated by a prime, becomes negative; that is, $m_{H'}^{2}<0$. It follows that the up and down neutral Higgs fields develop non-vanishing VEVs given by
\begin{equation}
\langle H^{0} \rangle^{2}=- \frac{m_{H'}^{2}}{g_{Y}^{2}+g_{2}^{2}}, \quad \langle \bar{H}^{0} \rangle= \frac{1}{tan\beta}\langle H^{0} \rangle \ .
\label{snow5}
\end{equation}
Evaluated at  $\langle H^{0} \rangle$, $ \langle \bar{H}^{0} \rangle$ and $\langle \nu_{3} \rangle$, all other VEVs vanish. Therefore, this is a stable vacuum which, while continuing to break $B$-$L$ symmetry at $\sim TeV$, now spontaneously breaks $SU(2)_{L} \times U(1)_{Y}$ to 
$U(1)_{EM}$ at the electroweak scale. This gives the $Z$ and $W^{\pm}$ vector bosons mass. Note that in our range of $tan\beta$, $  \langle \bar{H}^{0} \rangle\ll \langle H^{0} \rangle$. Hence, although included in the numerical analysis, to simplify equations we will not display any
$\langle \bar{H}^{0} \rangle\ $ contributions. For example, to leading order
\begin{equation}
M_{Z}=\sqrt{2}(g_{Y}^{2}+g_{2}^{2})^{1/2}\langle H^{0} \rangle \ .
 \label{snow6}
 \end{equation}
 The expressions for the slepton/squark masses at this minimum are each modified by additional terms proportional to the Higgs VEVs. By far the largest such contribution is to the third family left- and right-handed up squark masses from their Yukawa interaction in \eqref{6}. Ignoring much 
 smaller $D$-term corrections, these are given by
\begin{equation}
\langle \langle m_{U_{3}}^{2} \rangle \rangle = {\bf m_{U_{3}}^{2}} +\frac{1}{3} g_{B-L}^{2} \langle \nu_{3} \rangle^{2},\quad  \langle \langle m_{u_{3}}^{2} \rangle \rangle = {\bf m_{u_{3}}^{2}} -\frac{1}{3} g_{B-L}^{2} \langle \nu_{3} \rangle^{2} 
\label{snow7a}
\end{equation}
where
\begin{equation}
{\bf m_{U_{3}, u_{3}}^{2}} =  m_{Q_{3},u_{3}}^{2} +|\lambda_{u_{3}}|^{2}
\langle H^{0} \rangle^{2} \ .
\label{snow7b}
\end{equation}
Comparing to \eqref{snow4}, one sees that the mass parameters of $U_{3}$ and $u_{3}$ are modified by a positive Higgs VEV contribution. For this to be a stable vacuum, ${\bf m_{u_{3}}^{2}} $ must be positive at the electroweak scale. On the other hand, as long as $\langle \langle m_{U_{3}}^{2} \rangle \rangle$ is positive one can have ${\bf m_{U_{3}}^{2}}<0$. This is consistent with the conclusions at the $B$-$L$ scale. However, to leading order
\begin{equation}
\langle \langle m_{D_{3}}^{2} \rangle \rangle = m_{Q_{3}}^{2} +\frac{1}{3} g_{B-L}^{2} \langle \nu_{3} \rangle^{2} \ .
\label{snowc}
\end{equation}
It follows that if $m_{Q_{3}}^{2}<0$, the potential is most destabilized in the $D_{3}$ direction.
Note that all other Higgs VEV contributions, either through $F$-terms or $D$-terms, are much smaller. Hence, with the exception of the splitting of the $U_{3}$ and $D_{3}$ mass parameters, all conclusions regarding soft masses reached at the $B$-$L$ scale remain valid. For simplicity, we will {\it no longer notationally distinguish} between soft mass parameters $m^{2}$ and their Higgs corrected values ${\bf m^{2}}$.

Finally, using the above results one can calculate the $B$-$L$/electroweak hierarchy. It follows from \eqref{snow1} and \eqref{snow5} that
\begin{equation}
\frac{ \langle \nu_{3} \rangle}{\langle H^{0} \rangle} =\frac{\sqrt{g_{Y}^{2}+g_{2}^{2}}}{g_{B-L}} ~\frac{|m_{\nu_{3}}|}{|m_{H'}|} \ .
\label{snow9}
\end{equation}
In the analysis of~\cite{mike1,mike2}, $tan\beta$ was limited to $6.32 \leq tan\beta \leq 40$ and there was a specific range of initial parameters. For a generic choice in this range, it was found that
\begin{equation}
19.9 \leq \frac{ \langle \nu_{3} \rangle}{\langle H^{0} \rangle} \leq 126 \ .
\label{snow10}
\end{equation}
This demonstrates that a stable vacuum exists with an phenomenologically viable $B$-$L$/electroweak hierarchy. Within this range of parameters, $m_{\nu_{3}}^{2}<0$. All other slepton/squark soft masses are positive with the exception of $m_{Q_{3}}^{2}$, which is negative. Hence, although the $B$-$L$/electroweak vacuum is a minumum, {\it at the origin of field space} the potential is unstable in the $D_{3}$ direction. This has interesting applications to bosonic superconductivity and will be discussed in detail in Section 5. 

Recently, this analysis has been expanded to a {\it much larger} initial parameter space using a completely numerical calculation of the RGE's. This will appear elsewhere. Suffice it here to say that, over this entire extended range, masses  $m_{\nu_{3}}^{2}$ and $m_{H'}^{2}$ are negative at the electroweak scale and induce a viable hierarchy of
\begin{equation}
{\cal{O}}(10) \leq \frac{ \langle \nu_{3} \rangle}{\langle H^{0} \rangle} \leq {\cal{O}}(10^{2})  \ .
\label{snow10a}
\end{equation}
However, within this expanded context, the squark/slepton masses are considerably less constrained. Specifically, each of the following combinations of soft scalar mass parameters at the electroweak scale can now occur: 1) all positive, 2) all positive except for $m_{e_{3}}^{2}<0$, 3) all positive except for $m_{Q_{3}}^{2}<0$ and 4) combinations of these. We emphasize that in all cases the $B$-$L$/electroweak vacuum is a stable absolute minimum of the potential and does not break color or charge symmetry.

\section{The $B$-$L$ Cosmic String}

We begin by analyzing the theory at the $B$-$L$ breaking scale. The preceding results show that this symmetry is radiatively broken by a VEV of the third right-handed sneutrino. Furthermore, evaluated at this vacuum, all squark, slepton and Higgs mass squares  are positive. That is, this is a minimum of the potential energy and neither electroweak symmetry nor color is spontaneously broken at this scale. The situation at the origin of field space is more complex. As discussed above, it is possible for one or both of $m_{e_{3}}^{2}$ and $m_{Q_{3}}^{2}$ to be negative. However, to introduce the {\it basic} cosmic string solution, in this section we analyze the theory assuming {\it all} soft mass parameters are {\it positive}.

Under this assumption, the relevant physics is described by
\begin{equation}
{\cal{L}}=|{\cal{D}}_{\nu_{3}\mu} \nu_{3}|^{2} -\frac{1}{4}F_{B-L \mu\nu}F_{B-L}^{\mu\nu}-V(\nu_{3}) \ ,
\label{buddy1}
\end{equation}
where 
\begin{equation}
{\cal{D}}_{\nu_{3}\mu}=\partial_{\mu}-i g_{B-L}A_{B-L \mu}
\label{buddy1a}
\end{equation}
and
\begin{equation}
V(\nu_{3})=m_{\nu_{3}}^{2}|\nu_{3}|^{2}+\frac{g_{B-L}^{2}}{2}|\nu_{3}|^{4} \ .
\label{buddy2}
\end{equation}
The potential arises from two sources. The first term is the soft supersymmetry breaking third sneutrino mass term in \eqref{12} at the $B$-$L$ scale. The second term arises as the pure third sneutrino part of the $D_{B-L}$ supersymmetric contribution in \eqref{10}. Recall from the preceding RGE analysis that $m_{\nu_{3}}^{2}=-|m_{\nu_{3}}^{2}|$ at the $B$-$L$ scale. Hence, this potential is unstable at the origin and has a minimum at
\begin{equation}
\langle \nu_{3} \rangle^{2}=-\frac{m_{\nu_{3}}^{2}}{g_{B-L}^{2}} \ .
\label{buddy3}
\end{equation}
Using this, potential \eqref{buddy2} can be rewritten as
\begin{equation}
V(\nu_{3})=\frac {g_{B-L}^{2}}{2}(|\nu_{3}|^{2}-\langle\nu_{3}\rangle^{2})^{2} \ .
\label{buddy4}
\end{equation}
Note that the soft supersymmetry breaking $\nu_{3}$ mass term has been re-expressed as the Fayet-Iliopoulos component of an effective D-term. It follows from this that
the Higgs effect associated with \eqref{buddy3} gives the $A_{B-L}$ vector boson and the radial real scalar $\delta\nu_{3}$ an identical  mass
\begin{equation}
M_{A_{B-L}}=m_{\delta\nu_{3}}=\sqrt{2}g_{B-L} \langle\nu_{3}\rangle \ .
\label{newbuddy4}
\end{equation}

The cosmic string solution to this theory is well-known~\cite{Olsen}. Assuming a static solution that is translationally invariant in the $z$-coordinate, the cylindrically symmetric solution is of the form
\begin{equation}
{\bf \nu_{3}}=e^{i n \theta}\langle\nu_{3}\rangle f(r) \ , \qquad {\bf A_{B-Lr}}=0, \ {\bf A_{B-L \theta}}=\frac{n}{g_{B-L}r}\alpha(r) \ .
\label{buddy5}
\end{equation}
Here, integer $n$ is the ``winding number'' of the string around the origin, which will always be assumed non-zero. The functions $f(r)$ and $\alpha(r)$ have the boundary conditions
\begin{equation}
f\stackrel{r\rightarrow \infty}{\longrightarrow}1, \  f\stackrel{r\rightarrow 0}{\longrightarrow} 0  \qquad {\rm and} \qquad  
\alpha \stackrel{r\rightarrow \infty}{\longrightarrow}1,  \ \alpha \stackrel{r\rightarrow 0}{\longrightarrow} 0 
\label{buddy6}
\end{equation}
respectively. Before analyzing these functions further, it is important to note that there are two characteristic lengths associated with any cosmic string solution. These are
\begin{equation}
r_{s}=m_{\delta\nu_{3}}^{-1} \ , \qquad r_{v}=M_{A_{B-L}}^{-1} \ .
\label{buddy7}
\end{equation}
The explicit solutions for functions $f(r)$ and $\alpha(r)$ will depend on the ratio
\begin{equation}
{\cal{R}}=\frac{r_{v}^{2}}{r_{s}^{2}} \ .
\label{buddy8}
\end{equation}
In our case, we see from \eqref{newbuddy4} that
\begin{equation}
r_{s}=r_{v}=\frac{\langle \nu_{3} \rangle^{-1}}{\sqrt{2}g_{B-L}} \
\label{buddy9}
\end{equation}
and, hence, ${\cal{R}}$ is at the {\it critical point} 
\begin{equation}
{\cal{R}}=1 \ .
\label{buddy 10}
\end{equation}
This is a consequence of the softly broken supersymmetry of our theory, and will have important implications when we study bosonic superconductivity.
At the critical point, the equations for $f(r)$ and $\alpha(r)$ simplify to
\begin{equation}
f'=\frac{nf}{r}(1-\alpha) \ , \qquad \frac{\alpha'}{r}=\frac{1}{|n|} \langle \nu_{3} \rangle^{2}(f^{2}-1) 
\label{buddy11}
\end{equation}
where $ \ '  \ $ is the derivative with respect to $r$. Explicit solutions, even to these simplified equations, are not known, although their asymptotic expressions at small and large $r$ have been evaluated~\cite{Olsen}. However, precise numerical solutions for $f(r)$ and $\alpha(r)$ exist in the literature, See, for example,~\cite{Vilenkin}. We use both the asymptotic and numerical results throughout this paper. Another consequence of being at the critical point is that the energy density of the cosmic string simplifies to the exact result
\begin{equation}
\rho=2\pi \langle \nu_{3} \rangle^{2} \ .
\label{finalfinal}
\end{equation}

Let us now consider the theory at the electroweak breaking scale. As discussed in the previous section, 
the up and down neutral Higgs fields develop non-vanishing VEVs given by
\begin{equation}
\langle H^{0} \rangle^{2}=- \frac{m_{H'}^{2}}{g_{Y}^{2}+g_{2}^{2}}, \quad \langle \bar{H}^{0} \rangle= \frac{1}{tan\beta}\langle H^{0} \rangle \ .
\label{buddy12}
\end{equation}
Evaluated at  $\langle H^{0} \rangle$, $ \langle \bar{H}^{0} \rangle$ and $\langle \nu_{3} \rangle$, all other VEVs vanish. Therefore, this is a stable vacuum which breaks both $B$-$L$ and electroweak symmetries with a viable hierarchy.
Does the electroweak phase transition effect the basic $B$-$L$ cosmic string solution? Since both Higgs field have vanishing $B$-$L$ charge, $\nu_{3}$ is electroweak neutral and the Yukawa coupling $\lambda_{\nu_{3}}$ in \eqref{6} is of order $10^{-9}$, the Higgs VEV contribution to the dynamical equations for $\nu_{3}$ is highly suppressed. Furthermore, it remains possible to choose initial parameters so that all soft masses are positive, even at the  electroweak scale. It follows that the form of the cosmic string solution given in \eqref{buddy5} does not change. Although the Higgs VEVs are no longer zero, since these fields are $B$-$L$ neutral there are no topologically non--trivial solutions to the Higgs equations of motion~\cite{KibbleReview}. Henceforth, we assume that the Higgs fields are everywhere constants with the values given in \eqref{buddy12}.

\section{Bosonic Superconductivity}

Bosonic superconductivity can occur if a charged scalar field develops a non-vanishing condensate in the core of the cosmic string~\cite{Witten}. In the phenomenological $B$-$L$ MSSM theory discussed in this paper, there are a number of different ways that this could occur, each intricately related to other particle physics phenomena. Clearly, the first requirement for the existence of any such condensate is that a charged scalar mass squared at the origin of field space, that is, a soft supersymmetry breaking mass parameter plus small Higgs VEV corrections, becomes negative. As discussed in Section 3, there are several distinct ways in which this can occur. In this section, we examine bosonic superconductivity within the core of the cosmic string in each of these scenarios. 
The entire analysis will be carried out at the electroweak scale.

\subsection*{Case 1: All Soft Masses Positive}

This is the case described in the previous section. Since all soft supersymmetry breaking masses are positive, there can be no scalar condensates and, hence, no bosonic superconductivity at the core of the cosmic string. However, such strings could exhibit fermionic superconductivity. This will be discussed in Section 6.

\subsection*{Case 2: Negative Soft Slepton Mass}

As discussed in Section 3, there is a region of initial parameter space such that, at the electroweak scale, all soft masses are positive with the exception of $m_{e_{3}}^{2}<0$. This is the simplest case potentially admitting a non-zero condensate and, hence, we analyze it first.
The relevant Lagrangian for discussing the vacuum of $\nu_{3}$ and $e_{3}$ is given by
\begin{equation}
{\cal{L}}=|{\cal{D}}_{\nu_{3}\mu} \nu_{3}|^{2} -\frac{1}{4}F_{B-L \mu\nu}F_{B-L}^{\mu\nu}
+|{\cal{D}}_{e_{3}\mu}e_{3}|^{2} -\frac{1}{4}F_{Y \mu\nu}F_{Y}^{\mu\nu} -V(\nu_{3},e_{3})
\label{tues1}
\end{equation}
where
\begin{equation}
{\cal{D}}_{\nu_{3}\mu}= \partial_{\mu}-i g_{B-L}A_{B-L \mu} , \quad
{\cal{D}}_{e_{3}\mu}=\partial_{\mu}-i g_{B-L}A_{B-L \mu}-i g_{Y}A_{Y \mu}
\label{tues2}
\end{equation}
and 
\begin{equation} 
V(\nu_{3},e_{3})=m_{\nu_{3}}^{2}|\nu_{3}|^{2}+m_{e_{3}}^{2}|e_{3}|^{2}+\frac{g_{B-L}^{2}}{2}(|\nu_{3}|^{2}+|e_{3}|^{2})^{2} +\frac{g_{Y}^{2}}{2}|e_{3}|^{4} \ .
\label{tues3}
\end{equation}
The first two terms in the potential are the soft supersymmetry breaking mass terms in \eqref{12}, while the third and fourth terms are supersymmetric and arise from the $D_{B-L}$ and $D_{Y}$ in \eqref{10} and \eqref{9} respectively. Contributions to \eqref{tues3} from the relevant Yukawa couplings in \eqref{6} are suppressed, since $\lambda_{\nu_{3}}$ and $\lambda_{e_{3}}$ are of order $10^{-9}$ and $10^{-2}$ respectively. Hence, we ignore them. The RG analysis tells us that both $m_{\nu_{3}}^{2}<0, m_{e_{3}}^{2}<0$ at the electroweak scale. Hence, the potential is unstable at the origin of field space and has two other local extrema at
\begin{equation}
\langle \nu_{3} \rangle^{2}=-\frac{m_{\nu_{3}}^{2}}{g_{B-L}^{2}} , \quad \langle e_{3} \rangle=0
\label{tues4}
\end{equation}
and
\begin{equation}
\langle \nu_{3}\rangle= 0, \quad \langle e_{3}\rangle^{2}=-\frac{m_{e_{3}}^{2}}{g_{B-L}^{2}+g_{Y}^{2}} 
\label{tues5}
\end{equation}
respectively. Using these, potential \eqref{tues3} can be rewritten as
\begin{eqnarray} 
&& V(\nu_{3},e_{3})= \frac {g_{B-L}^{2}}{2}(|\nu_{3}|^{2}-\langle \nu_{3}\rangle^{2})^{2} +g_{B-L}^{2} |\nu_{3}|^{2} |e_{3}|^{2} \nonumber \\
&& \qquad \qquad +\frac {g_{B-L}^{2} 
+g_{Y}^{2}}{2}(|e_{3}|^{2}-\langle e_{3}\rangle^{2})^{2} \ .
\label{tues6}
\end{eqnarray}

Let us analyze these two extrema. Both have positive masses in their radial directions. At the sneutrino vacuum \eqref{tues4}, the mass squared in the $e_{3}$ direction is given by
\begin{equation}
m_{e_{e}}^{2}|_{\langle \nu_{3}\rangle}= g_{B-L}^{2}\langle \nu_{3}\rangle^{2}-(g_{B-L}^{2}+g_{Y}^{2})\langle e_{3}\rangle^{2}= |m_{\nu_{3}}|^{2}-|m_{e_{3}}|^{2} \ ,
\label{tues7}
\end{equation}
whereas at the stau vacuum \eqref{tues5}, the mass squared in the $\nu_{3}$ direction is 
\begin{equation}
m_{\nu_{3}}^{2}|_{\langle e_{3}\rangle}= g_{B-L}^{2}\langle e_{3}\rangle^{2}-g_{B-L}^{2} \langle \nu_{3}\rangle^{2}= |m_{e_{3}}|^{2}(1+\frac{g_{Y}^{2}}{g_{B-L}^{2}})^{-1}-|m_{\nu_{3}}|^{2} \ .
\label{tues8}
\end{equation}
Note that either \eqref{tues7} or \eqref{tues8} can be positive, but not both. To be consistent with the hierarchy solution, we want \eqref{tues4} to be a stable minimum. Hence, we demand $m_{e_{3}}^{2}|_{\langle \nu_{3}\rangle}>0$ or, equivalently, that
\begin{equation}
|m_{\nu_{3}}|^{2}>|m_{e_{3}}|^{2} \ .
\label{tues9}
\end{equation}
It follows from the RG analysis in~\cite{mike3} that one can always find a subregion of the initial parameter space so that this condition holds. We assume \eqref{tues9} for the remainder of this subsection. It then follows from \eqref{tues8} that $m_{\nu_{3}}^{2}|_{\langle e_{3}\rangle}<0$ and, hence, the stau extremum \eqref{tues5} is a saddle point. As a consistency check, note that $V|_{\langle\nu_{3}\rangle}<V|_{\langle e_{3}\rangle}$ if and only if 
\begin{equation}
g_{B-L}^{2}\langle \nu_{3}\rangle^{4}>(g_{B-L}^{2}+g_{Y}^{2})\langle e_{3}\rangle^{4} 
\label{tues10}
\end{equation}
or, equivalently, 
\begin{equation}
|m_{\nu_{3}}|^{2}>|m_{e_{3}}|^{2}(1+\frac{g_{Y}^{2}}{g_{B-L}^{2}})^{-1/2} \ .
\label{tues11}
\end{equation}
This follows immediately from constraint \eqref{tues9}. Finally, note that the potential descends monotonically along a path  ${\cal{C}}$ from the saddle point at \eqref{tues5} to the absolute minimum at \eqref{tues4}. Solving the $\frac{\partial V}{\partial {e_{3}}}=0$ equation, this curve is found to be
\begin{equation}
|e_{3}|_{{\cal{C}}}=( \langle e_{3}\rangle^{2}-|\nu_{3}|^{2}(1+\frac{g_{Y}^{2}}{g_{B-L}^{2}})^{-1})^{1/2} \ .
\label{tues12}
\end{equation}
Note that it begins at $ \langle e_{3}\rangle$ for $\nu_{3}=0$ and continues until it tangentially intersects the $e_{3}=0$ axis at $|\nu_{30}|=\frac{|m_{e_{3}}|}{|m_{\nu_{3}}|}\langle \nu_{3}\rangle$. From here, the path continues down this axis to the stable minimum at \eqref{tues4}.

We conclude that at the electroweak scale the absolute minimum of potential \eqref{tues3} occurs at the sneutrino vacuum given in \eqref{tues4}. The sneutrino scalar can develop a non-zero winding around some point in three-space, leading to a cosmic string. Away from the core, this will still be described by the simple cosmic string solution in the previous section. 
Recall, however, that non-zero winding forces the function $f(r)$ and, hence, ${\bf \nu_{3}}$ to vanish at $r=0$.  This was not an issue for the simple cosmic string, since it was assumed that all squark/slepton masses were positive at the origin of field space. In the 
present scenario, however, the mass squared of $e_{3}$, 
\begin{equation}
m_{e_{3}}^{2}|_{\bf \nu_{3} }= m_{e_{3}}^{2}+g_{B-L}^{2}{\bf \nu_{3}}^{2} \ ,
\label{tues12a}
\end{equation}
becomes negative as ${\bf \nu_{3}}$ approaches the origin of field space. This potentially destabilizes the $e_{3}$ field in the core of the string, producing a scalar condensate. Whether or not this can occur is dependent on the relative magnitudes of the spatial gradient and the potential energy, which tend to stabilize and destabilize $e_{3}$ respectively. To analyze this, one can look at small fluctuations of $e_{3}$ around zero in the background of the simple $\nu_{3}$ cosmic string solution in Section 4. The equation of motion for $e_{3}$ is given, to linear order, by
\begin{eqnarray}
&& (\partial_{\mu} \partial^{\mu} +2ig_{B-L}{\bf A_{B-L\theta}}\partial_{\theta}-g_{B-L}^{2}{\bf A_{B-L\mu}}{\bf A_{B-L}^{\mu}}) e_{3} \nonumber \\
&& \qquad + (g_{B-L}^{2}|{\bf \nu_{3}}|^{2}-(g_{B-L}^{2}+g_{Y}^{2})\langle e_{3}\rangle^{2}) e_{3}= 0 \ ,
\label{tues13}
\end{eqnarray}
where ${\bf \nu_{3}}$ and ${\bf A_{B-L\mu}}$ were defined in \eqref{buddy5}. Using the ansatz 
\begin{equation}
{\bf e_{3}}=e^{i\omega t} {\bf e_{3}}_{0}(r) \ ,
\label{tues14}
\end{equation}
equation \eqref{tues13} simplifies to
\begin{equation}
( -\frac{\partial^{2}}{\partial r^{2}} -\frac{1}{r}\frac{\partial}{\partial r}){\bf e_{3}}_{0}+ {\hat{V}}{\bf e_{3}}_{0}=\omega^{2}{\bf e_{3}}_{0}
\label{tues15}
\end{equation}
where
\begin{equation}
{\hat{V}}(r)=g_{B-L}^{2}\langle \nu_{3}\rangle^{2}f(r)^{2}-(g_{B-L}^{2}+g_{Y}^{2}) \langle e_{3}\rangle^{2}+n^{2}\frac{\alpha(r)^{2}}{r^{2}} \ .
\label{tues16}
\end{equation}
Note that we have chosen ${\bf e_{3}}_{0}$ in \eqref{tues14} to be a function of radial coordinate $r$ only and, hence, not to wind around the origin. If this two-dimensional Sturm-Liouville equation has at least one negative eigenvalue, the corresponding $\omega$ becomes imaginary. This destabilizes ${\bf e_{3}}$, implying the existence of an $e_{3}$ condensate in the core of the cosmic string.

\subsection*{Case 3: Negative Soft Squark Mass}

As discussed in Section 3, there is a region of initial parameter space such that, at the electroweak scale, all soft masses are positive with the exception of $m_{Q_{3}}^{2}<0$. The electroweak phase transition breaks the left-handed $SU(2)_{L}$ doublet $Q_{3}$ into its up- and down- quark components $U_{3}$ and $D_{3}$ respectively. The leading order contribution of the Higgs VEVs to their mass 
splits the degeneracy between these two fields, destabilizing the potential most strongly in the $D_{3}$ direction. For this reason, the relevant Lagrangian for analyzing this vacuum can be restricted to
\begin{eqnarray}
&& {\cal{L}}=|{\cal{D}}_{\nu_{3}\mu} \nu_{3}|^{2} -\frac{1}{4}F_{B-L \mu\nu}F_{B-L}^{\mu\nu}
+|{\cal{D}}_{D_{3}\mu}D_{3}|^{2} -\frac{1}{4}F_{Y \mu\nu}F_{Y}^{\mu\nu} \nonumber \\
&& \quad \  \  \ -\frac{1}{4}F_{SU(2) \mu\nu}F_{SU(2)}^{\mu\nu}-\frac{1}{4}F_{SU(3) \mu\nu}F_{SU(3)}^{\mu\nu} 
-V(\nu_{3},D_{3})
\label{wo1}
\end{eqnarray}
where
\begin{eqnarray}
&& {\cal{D}}_{\nu_{3}\mu}= \partial_{\mu}-i g_{B-L}A_{B-L \mu} \ , \label{wo2}  \\
&& {\cal{D}}_{D_{3}\mu}=\partial_{\mu}-i \frac{g_{B-L}}{3} A_{B-L \mu}-i \frac{g_{Y}}{6}A_{Y \mu}-i g_{2}A_{SU(2) \mu}-i g_{3}A_{SU(3) \mu} \nonumber
\end{eqnarray}
and 
\begin{eqnarray} 
&& V(\nu_{3},D_{3})=m_{\nu_{3}}^{2}|\nu_{3}|^{2}+m_{D_{3}}^{2}|D_{3}|^{2}+\frac{g_{B-L}^{2}}{2}(|\nu_{3}|^{2}+\frac{1}{3}|D_{3}|^{2})^{2} \label{wo3} \\  
&& \qquad \qquad \quad \ +\frac{1}{2}( \frac{g_{Y}^{2}}{36} +\frac{g_{2}^{2}}{4} +\frac{g_{3}^{2}}{3})|D_{3}|^{4} \ . \nonumber
\end{eqnarray}
The first two terms in the potential are the soft supersymmetry breaking mass terms in \eqref{12}, while the remaining terms are supersymmetric and arise from the $D_{B-L}$,  $D_{Y}$ in \eqref{10},\eqref{9} and $D_{SU(2)_{L}}$, $D_{SU(3)_{C}}$ respectively. 
Using $\lambda_{d_{3}} \simeq 5 \times 10^{-2}$, 
the hierarchy given in \eqref{snow10a} 
and assuming $|m_{D_{3}}|$ is of order $|m_{\nu_{3}}|$, terms proportional to the Higgs VEVs are small and are ignored in \eqref{wo3}. For simplicity, we henceforth drop the small $g_{B-L}^{2}/9+g_{Y}^{2}/36$ piece of the $D$-term contribution.
The RG analysis tells us that both $m_{\nu_{3}}^{2}<0, m_{D_{3}}^{2}<0$ at the electroweak scale.
Hence, the potential is unstable at the origin of field space and has two other local extrema at
\begin{equation}
\langle \nu_{3} \rangle^{2}=-\frac{m_{\nu_{3}}^{2}}{g_{B-L}^{2}} , \quad \langle D_{3} \rangle=0 \ ,
\label{wo4}
\end{equation}
and
\begin{equation}
\langle \nu_{3}\rangle= 0, \quad \langle D_{3}\rangle^{2} = -\frac{m_{D_{3}}^{2}}{g_{2}^{2}/4+g_{3}^{2}/3}
\label{wo5}
\end{equation}
respectively. Using these, potential \eqref{wo3} can be rewritten as
\begin{eqnarray} 
&& V(\nu_{3},D_{3})= \frac {g_{B-L}^{2}}{2}(|\nu_{3}|^{2}-\langle \nu_{3}\rangle^{2})^{2} +\frac{g_{B-L}^{2}}{3} |\nu_{3}|^{2} |D_{3}|^{2} \nonumber \\
&& \qquad \qquad \quad +\frac {g_{2}^{2}/4+g_{3}^{2}/3}{2}(|D_{3}|^{2}-\langle D_{3}\rangle^{2})^{2} \ .
\label{wo6}
\end{eqnarray}

Let us analyze these two extrema. Both have positive masses in their radial directions. At the sneutrino vacuum \eqref{wo4}, the mass squared in the $D_{3}$ direction is given by
\begin{equation}
m_{D_{3}}^{2}|_{\langle \nu_{3}\rangle}= \frac{g_{B-L}^{2}}{3}\langle \nu_{3}\rangle^{2}-(\frac{g_{2}^{2}}{4}+\frac{g_{3}^{2}}{3}) \langle D_{3}\rangle^{2}= \frac{|m_{\nu_{3}}|^{2}}{3}-|m_{D_{3}}|^{2} \ ,
\label{wo7}
\end{equation}
whereas at the stau vacuum \eqref{wo5}, the mass squared in the $\nu_{3}$ direction is 
\begin{equation}
m_{\nu_{3}}^{2}|_{\langle D_{3}\rangle}= \frac{g_{B-L}^{2}}{3}\langle D_{3}\rangle^{2}-g_{B-L}^{2} \langle \nu_{3}\rangle^{2}=|m_{D_{3}}|^{2} (\frac{g_{B-L}^{2}}{3g_{2}^{2}/4+g_{3}^{2}})-|m_{\nu_{3}}|^{2} \ .
\label{wo8}
\end{equation}
Note that either \eqref{wo7} or \eqref{wo8} can be positive, but not both. To be consistent with the hierarchy solution, we want \eqref{wo4} to be a stable minimum. Hence, we demand $m_{D_{3}}^{2}|_{\langle \nu_{3}\rangle}>0$ or, equivalently, that
\begin{equation}
|m_{\nu_{3}}|^{2}>3|m_{D_{3}}|^{2} \ .
\label{wo9}
\end{equation}
The RG analysis in~\cite{mike2} shows that one can always find a region of the initial parameter space so that this condition holds. We assume \eqref{wo9} for the remainder of this subsection. It then follows from \eqref{wo8} that $m_{\nu_{3}}^{2}|_{\langle D_{3}\rangle}<0$ and, hence, the stau extremum \eqref{wo5} is a saddle point. As a consistency check, note that $V|_{\langle\nu_{3}\rangle}<V|_{\langle D_{3}\rangle}$ if and only if
\begin{equation}
g_{B-L}^{2}\langle \nu_{3}\rangle^{4}>(\frac{g_{2}^{2}}{4}+\frac{g_{3}^{2}}{3}) \langle D_{3}\rangle^{4} 
\label{wo10}
\end{equation}
or, equivalently, 
\begin{equation}
|m_{\nu_{3}}|^{2}>|m_{D_{3}}|^{2} (\frac{g_{B-L}^{2}}{3g_{2}^{2}/4+g_{3}^{2}})^{1/2} \ .
\label{wo11}
\end{equation}
This follows immediately from constraint \eqref{wo9}. 

We conclude that at the electroweak scale the absolute minimum of potential \eqref{wo3} occurs at the sneutrino vacuum given in \eqref{wo4}. The sneutrino scalar can develop a non-zero winding around some point in three-space, leading to a cosmic string. This is described, away from the core, by the simple cosmic string solution in the previous section.  
Recall, however, that non-zero winding forces the the function $f(r)$ and, hence, ${\bf \nu_{3}}$ to vanish at $r=0$. In the present scenario, 
the mass squared of $D_{3}$, 
\begin{equation}
m_{D_{3}}^{2}|_{\bf \nu_{3} }= m_{D_{3}}^{2}+\frac{g_{B-L}^{2}}{3}{\bf \nu_{3}}^{2} \ ,
\label{wo11a}
\end{equation}
becomes negative as ${\bf \nu_{3}}$ approaches the origin of field space. This potentially destabilizes the $D_{3}$ field in the core of the string, producing a scalar condensate. 
To analyze this, one can look at small fluctuations of $D_{3}$ around zero in the background of the simple $\nu_{3}$ cosmic string solution. The equation of motion for $D_{3}$ is given, to linear order, by
\begin{eqnarray}
&& (\partial_{\mu} \partial^{\mu} +2i\frac{g_{B-L}}{3}{\bf A_{B-L\theta}}\partial_{\theta}-\frac{g_{B-L}^{2}}{9}{\bf A_{B-L\mu}}{\bf A_{B-L}^{\mu}}) D_{3} \nonumber \\
&& \qquad + (\frac{g_{B-L}^{2}}{3}|{\bf \nu_{3}}|^{2}-(\frac{g_{2}^{2}}{4}+\frac{g_{3}^{2}}{3})\langle D_{3}\rangle^{2}) D_{3}= 0 \ ,
\label{wo13}
\end{eqnarray}
where ${\bf \nu_{3}}$ and ${\bf A_{B-L\mu}}$ were defined in \eqref{buddy5}. Using the ansatz 
\begin{equation}
{\bf D_{3}}=e^{i\omega t} {\bf D_{3}}_{0}(r) \ ,
\label{wo14}
\end{equation}
equation \eqref{wo13} simplifies to
\begin{equation}
( -\frac{\partial^{2}}{\partial r^{2}} -\frac{1}{r}\frac{\partial}{\partial r}){\bf D_{3}}_{0}+ {\hat{V}}{\bf D_{3}}_{0}=\omega^{2}{\bf D_{3}}_{0}
\label{wo15}
\end{equation}
where
\begin{equation}
{\hat{V}}(r)=\frac{g_{B-L}^{2}}{3}\langle \nu_{3}\rangle^{2}f(r)^{2}-(\frac{g_{2}^{2}}{4}+\frac{g_{3}^{2}}{3}) \langle D_{3}\rangle^{2}+\frac{n^{2}}{9}\frac{\alpha(r)^{2}}{r^{2}} \ .
\label{wo16}
\end{equation}
Note that we have chosen ${\bf D_{3}}_{0}$ in \eqref{wo14} to be a function of radial coordinate $r$ only and, hence, not to wind around the origin. If this two-dimensional Sturm-Liouville equation has at least one negative eigenvalue, the corresponding $\omega$ becomes imaginary. This destabilizes ${\bf D_{3}}$, implying the existence of an $D_{3}$ condensate in the core of the cosmic string.

\subsection*{Numerical Analysis of Boson Condensates}

Let us analyze the formation of a scalar condensate in a more general setting. Consider a $U(1) \times {\tilde{U}}(1)$ gauge theory with two complex scalar fields $\phi$ and $\sigma$ charged under the gauge group as $q_{\phi}=0$, $\tilde{q}_{\phi}\neq 0$ and $q_{\sigma} \neq0$, $\tilde{q}_{\sigma} \neq 0$ respectively. $U(1)$ and $\tilde{U}(1)$ are
motivated by $U_{Y}$ and $U_{B-L}$ in the previous sections. Similarly, scalar $\phi$ corresponds to the right-handed sneutrino $\nu_{3}$. A condensate can potentially form in the $\sigma$ field. Unlike previous analyses in the literature, here, in addition to the usual $U(1)$ charge of $\sigma$, $\tilde{q}_{\sigma}$ is also non-vanishing. This is motivated by the fact that all squarks and sleptons in the $B$-$L$ MSSM theory carry non-vanishing $B$-$L$ charge. After finding the necessary conditions for a condensate to form, we will apply the results to the specific cases discussed above. 
 
 The Lagrangian density for this generic theory is given by
\begin{equation}
{\cal{L}}=|\tilde{{{\cal{D}}}}_{\mu} \phi|^{2} -\frac{1}{4}\tilde{F}_{\mu\nu}\tilde{F}^{\mu\nu}
+|{\cal{D}}_{\mu} \sigma|^{2} -\frac{1}{4}F_{\mu\nu}F^{\mu\nu} -V(\phi,\sigma)
\label{tues1A}
\end{equation}
where
\begin{equation}
\tilde{{\cal{D}}}_{\mu}= \partial_{\mu}-i {\tilde{q}}_{\phi} {\tilde{g}}{\tilde{A}}_{\mu} , \quad
{\cal{D}}_{\mu}=\partial_{\mu}-i q_{\sigma}gA_{\mu}-i {\tilde{q}}_{\sigma}{\tilde{g}}{\tilde{A}}_{\mu}
\label{tues2A}
\end{equation}
and 
\begin{equation} 
V(\phi,\sigma)=\frac{\lambda_{\phi}}{4}(|\phi|^{2}-\eta_{\phi}^{2})^{2} +\beta|\phi|^{2}|\sigma|^{2}+\frac{\lambda_{\sigma}}{4}(|\sigma|^{2}-\eta_{\sigma}^{2})^{2}  \ .
\label{tues3A}
\end{equation}
The coefficients $\lambda_{\phi}$,$\lambda_{\sigma}$ and $\beta$ are chosen to be positive. Potential \eqref{tues3A} has an extremum at $\langle \phi \rangle=\eta_{\phi}$,~$\langle\sigma\rangle=0$. If one chooses the coefficients so that the effective $\sigma$ mass squared at this extremum is positive, that is,
\begin{equation}
m_{\sigma}^{2}|_{\eta_{\phi}}=\beta \eta_{\phi}^{2}-\frac{\lambda_{\sigma}\eta_{\sigma}}{2} > 0 \ ,
\label{mass1}
\end{equation}
then $\langle \phi \rangle=\eta_{\phi}$,~$\langle \sigma \rangle=0$ is a local minimum. This vacuum spontaneously breaks the $\tilde{U}(1)$ symmetry and admits a cosmic string solution in $\phi$ of the form discussed in Section 4. Potential \eqref{tues3A} has a second extremum at $\langle \phi \rangle=0$,~$\langle\sigma\rangle=\eta_{\sigma}$. This may or may not be a local minimum of the potential depending on the choice of parameters. In all cases, however, one can constrain the cosmic string vacuum to be deeper than the $\sigma$ extremum by choosing
\begin{equation}
\lambda_{\phi} \eta_{\phi}^{4} > \lambda_{\phi} \eta_{\phi}^{4} \ ,
\label{constr1}
\end{equation}
which we do henceforth.

As discussed in Section 4, there is a $\tilde{U}(1)$ cosmic string solution of the associated 
$\phi$ and $\tilde{A}_{\mu}$ equations of  motion given by
\begin{equation}
{\bf \phi}=e^{i n \theta}\eta_{\phi} f(r) \ , \qquad {\bf \tilde{A}_{r}}=0, \ {\bf \tilde{A}_{ \theta}}=\frac{n}
{\tilde{q}_{\phi}\tilde{g}r}\alpha(r) 
\label{buddy5A}
\end{equation}
where integer $n$ is the ``winding number'' of the string around the origin. The functions $f(r)$ and $\alpha(r)$ have the boundary conditions given in \eqref{buddy6}.
In the theory we are considering, the effective mass squared of $\sigma$ at arbitrary $\phi$ is 
\begin{equation}
m_{\sigma}^{2}|_{\bf \phi }=\beta|{\bf \phi}|^{2} -\frac{\lambda_{\sigma}\eta_{\sigma}^{2}}{2} \ .
\label{tues12aA}
\end{equation}
This becomes negative as ${\bf \phi}$ approaches the origin of field space, potentially destabilizing 
the $\sigma$ field in the core of the string and producing a scalar condensate. 
Such a condensate would break both $\tilde{U}(1)$ and $U(1)$ symmetry.
Whether or not this can occur is dependent on the relative magnitudes of the spatial gradient and the potential energy, which tend to stabilize and destabilize $\sigma$ respectively. To analyze this, one can look at small fluctuations of 
$\sigma$ around zero in the background of the simple $\bf \phi$ cosmic string solution. The equation of motion for $\sigma$ is given, to linear order, by
\begin{eqnarray}
&& (\partial_{\mu} \partial^{\mu} +2i\tilde{q}_{\sigma}\tilde{g}{\bf \tilde{A}_{\theta}}\partial_{\theta}-
\tilde{q}_{\sigma}^{2}\tilde{g}^{2}
{\bf \tilde{A}_{\mu}}{\bf \tilde{A}^{\mu}}) \sigma \nonumber \\
&& \qquad +(\beta |{\bf \phi}|^{2}-\frac{\lambda_{\sigma}\eta_{\sigma}^{2}}{2})\sigma= 0 \ ,
\label{tues13A}
\end{eqnarray}
where ${\bf \phi}$ and ${\bf \tilde{A}_{\mu}}$ were defined in \eqref{buddy5A}. Using the ansatz 
\begin{equation}
{\bf \sigma}=e^{i\omega t} {\bf \sigma}_{0}(r) \ ,
\label{tues14A}
\end{equation}
equation \eqref{tues13A} simplifies to
\begin{equation}
( -\frac{\partial^{2}}{\partial r^{2}} -\frac{1}{r}\frac{\partial}{\partial r}){\bf \sigma}_{0}+ {\hat{V}}{\bf \sigma}_{0}=\omega^{2}{\bf \sigma}_{0}
\label{tues15}
\end{equation}
where
\begin{equation}
{\hat{V}}(r)=\beta \eta_{\phi}^{2}f(r)^{2}-\frac{\lambda_{\sigma}\eta_{\sigma}^{2}}{2}+
n^{2}(\frac{\tilde{q}_{\sigma}^{2}}{\tilde{q}_{\phi}^{2}}) \frac{\alpha(r)^{2}}{r^{2}} \ .
\label{tues16}
\end{equation}
Note that we have chosen ${\bf \sigma}_{0}$ in \eqref{tues14A} to be a function of radial coordinate $r$ only and, hence, not to wind around the origin. We want to emphasize the term in \eqref{tues16} proportional to $\alpha^{2}/r^{2}$. This appears precisely because the $\sigma$ field has non-vanishing charge under $\tilde{U}(1)$ as well as under $U(1)$, a situation not previously discussed in the literature. However, for the reasons mentioned above, it must be included in the analysis of this paper.

If this two-dimensional Sturm-Liouville equation has at least one negative eigenvalue, the corresponding $\omega$ becomes imaginary. This destabilizes ${\bf \sigma}$, implying the existence of a $\sigma$ condensate in the core of the cosmic string. Note that if the condensate was not charged under $\tilde{U}(1)$, then the last term in \eqref{tues16} would not appear and potential ${\hat{V}}(r)$
would be monotonic. As was discussed in~\cite{Witten}, for sufficiently small $m_{\sigma}^{2}|_{\eta_{\phi}}$ a condensate will always form under these conditions. However, $\sigma$ {\it is} charged under $\tilde{U}(1)$ and, hence, the $\alpha^{2}/r^{2}$ term in \eqref{tues16} must be included. Since this term is positive and provides a repellent force for large $r$, it may prevent  a bound state from forming. For the remainder of this section, we will discuss the results of a numerical solution to the Sturm-Liouville equation \eqref{tues15} with potential \eqref{tues16}. The details of this solution are presented in the Appendix. 

The $\alpha^{2}/r^{2}$ term is smallest and, hence, the least disruptive to the formation of a condensate for winding number $n=1$. Therefore, we carry out the analysis for the singly wound cosmic string. Motivated by the softly broken supersymmetric $B$-$L$ MSSM theory, the calculation will be further restricted in two ways. First, take $\tilde{q}_{\sigma}^{2}=\tilde{q}_{\phi}^{2}$. Second, we constrain the parameters to the critical coupling point where 
\begin{equation}
\frac{\lambda_{\phi}}{2\tilde{g}^{2}}=1 \ .
\label{ny1}
\end{equation}
It follows that the functions $f(r)$ and $\alpha(r)$ simplify to solutions of \eqref{buddy11}. These equations can be solved  numerically~\cite{Vilenkin}, and we input them into our analysis of the Sturm-Liouville equation. Finally, simplification can be achieved if we take 
\begin{equation}
\beta \eta_{\phi}^{2}=\frac{\lambda_{\sigma}\eta_{\sigma}^{2}}{2} \ ,
\label{ny2}
\end{equation}
thus setting $m_{\sigma}^{2}$ at the cosmic string vacuum, given in \eqref{mass1}, to zero. Potential 
${\hat{V}}(r)$ then becomes
\begin{equation}
{\hat{V}}(r)=\beta \eta_{\phi}^{2}(f(r)^{2}-1)+\frac{\alpha(r)^{2}}{r^{2}} \ .
\label{tues17}
\end{equation}
Should a negative energy bound state exist for some choice of $\beta$, the condensate will persist if we continuously deform $m_{\sigma}^{2}|_{\eta_{\phi}}$ away from zero to a small positive value.

In units where $\eta_{\phi}$ is one, the Sturm-Liouville equation \eqref{tues15} with potential \eqref{tues17} depends on the single parameter $\beta$.  Explicit solutions of this equation for several values of $\beta$ are shown in Figure 1. For each of these values, a negative energy eigenvalue and normalizable bound state wavefunction exists and are shown in the Figure. As discussed in the Appendix, we find that a negative energy eigenvalue will exist for any
\begin{equation}
\beta > \beta_{critical} \simeq 0.42 \ .
\label{acel1}
\end{equation}
Hence, for sufficiently large $\beta$ satisfying \eqref{acel1}, that is, for sufficiently deep potential, the $\sigma$ field is destabilized and a non-vanishing $\sigma$ condensate will form. However, for $\beta$ less than $\beta_{critical}$ the eigenvalue becomes positive and the wavefunction oscillatory, signaling a meta-stable solution. Hence, for $\beta< 0.42$ the potential is not sufficiently deep and a $\sigma$ condensate will ${\it not}$ form. 
These results can immediately be applied to the $B$-$L$ MSSM theory with a negative soft right-handed slepton mass $m_{e_{3}}^{2}<0$ described in Case 2 
\begin{figure}[h!]
\centering
\includegraphics[trim=4cm 16.5cm 1.5cm 3cm clip]{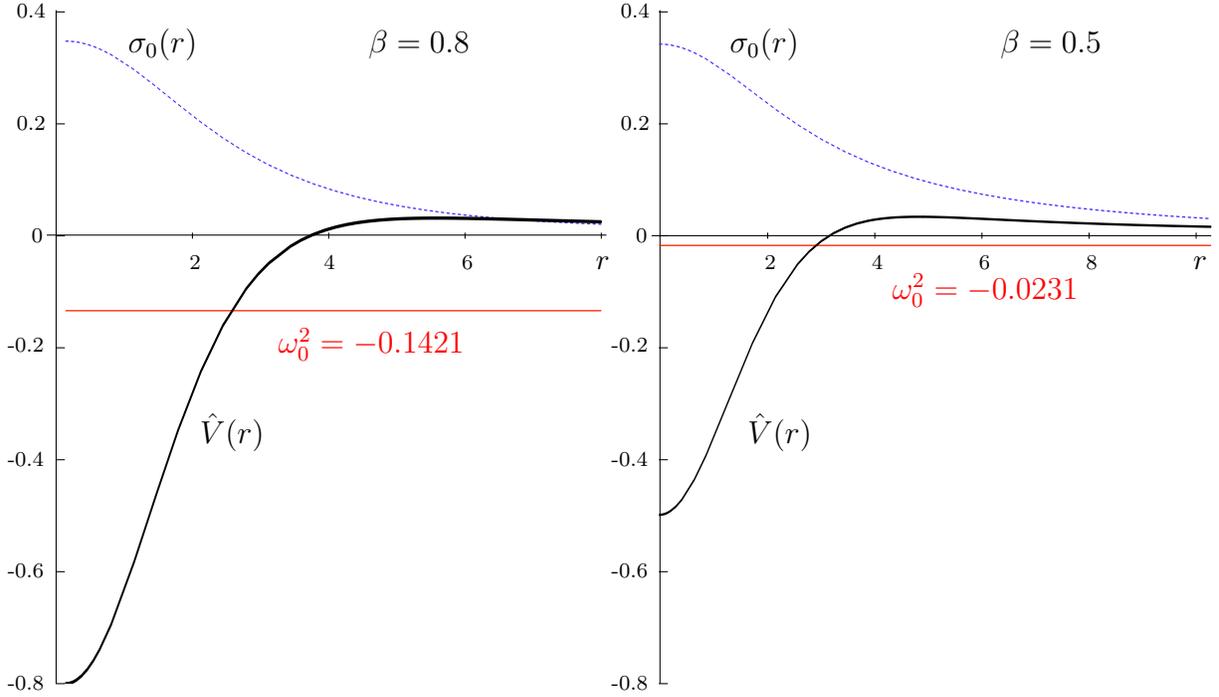}

\caption{Negative energy ground state solutions of the slepton stability equation for 
$\beta=0.8$ and $\beta=0.5$ respectively. The energy eigenvalues $\omega_{0}^{2}$ are shown as red lines, with the associated normalizable wave functions $\sigma_{0}$ depicted in blue. Note the positive ``bump'' in potential $\hat{V}$ due to the $\alpha^{2}/r^{2}$ term.}
\label{condensate}

\end{figure}
above. Identifying $\phi=\nu_{3}$, $\sigma=e_{3}$ and comparing \eqref{tues2A},\eqref{tues3A} to \eqref{tues2},\eqref{tues6} using \eqref{tues4},\eqref{tues5}, we find that
\begin{eqnarray}
&& \beta=g_{B-L}^{2} , \  \  \lambda_{\phi}=2g_{B-L}^{2} , \  \  \lambda_{\sigma}=2(g_{B-L}^{2}+g_{Y}^{2}) , \ \   \nonumber \\  
&& \qquad \eta_{\phi}^{2}=-\frac{m_{\nu_{3}}^{2}}{g_{B-L}^{2}}, \  \  \eta_{\sigma}^{2}=-\frac{m_{e_{3}}^{2}}{g_{B-L}^{2}+g_{Y}^{2}} \ .
\label{acel2}
\end{eqnarray}
In particular, evaluated at the electroweak scale
\begin{equation}
\beta=g_{B-L}^{2} \simeq 0.1075 < 0.42  
\label{acel3}
\end{equation}
suggesting the absense of a bosonic $e_{3}$ condensate in the core of the cosmic string. 

The formalism applicable to the $B$-$L$ MSSM theory with a negative soft left-handed squark mass squared requires a change in the relative charges of $\phi$ and $\sigma$.
Instead of taking $\tilde{q}_{\sigma}^{2}=\tilde{q}_{\phi}^{2}$ as we did previously, now choose
$\tilde{q}_{\phi}^{2}=9\tilde{q}_{\sigma}^{2}$. For $n=1$ winding number at the critical point and vanishing $m_{\sigma}^{2}|_{\eta_{\phi}}$, the Sturm-Liouville equation determining the $\sigma$
condensate is \eqref{tues15} with potential \eqref{tues16} now given by
\begin{equation}
{\hat{V}}(r)=\beta \eta_{\phi}^{2}(f(r)^{2}-1)+\frac{\alpha(r)^{2}}{9r^{2}} \ .
\label{acel4}
\end{equation}
A numerical analysis completely analogous to the one used above, leads to the conclusion that a non-vanishing $\sigma$ condensate will occur for 
\begin{equation}
\beta  > \beta_{critical} \simeq 0.14 \ .
\label{acel5}
\end{equation}
Note that this is smaller than the previous bound given in \eqref{acel1}. This is consistent with expectations since the destabilizing $\alpha^{2}/r^{2}$ term is now smaller by a factor of $9$.
For $\beta$ less than $\beta_{critical}$, however, the eigenvalue becomes positive and the wavefunction oscillatory, signaling a meta-stable solution. Hence, for $\beta< 0.14$ the potential is not sufficiently deep and a $\sigma$ condensate will ${\it not}$ form. 
By construction, these results can immediately be applied to the $B$-$L$ MSSM theory with a negative soft left-handed squark mass $m_{D_{3}}^{2}<0$ described in Case 3 above. Identifying $\phi=\nu_{3}$, $\sigma=D_{3}$ and comparing \eqref{tues2A},\eqref{tues3A} to \eqref{wo2},\eqref{wo6} using \eqref{wo4},\eqref{wo5}, we find that
\begin{eqnarray}
&& \beta=\frac{g_{B-L}^{2}}{3} , \  \  \lambda_{\phi}=2g_{B-L}^{2} , \  \  \lambda_{\sigma}=2(\frac{g_{2}^{2}}{4}+\frac{g_{3}^{2}}{3}) , \ \   \nonumber \\  
&& \qquad \eta_{\phi}^{2}=-\frac{m_{\nu_{3}}^{2}}{g_{B-L}^{2}}, \  \  \eta_{\sigma}^{2}=-\frac{m_{D_{3}}^{2}}{g_{2}^{2}/4+g_{3}^{2}/3} \ .
\label{acel2}
\end{eqnarray}
In particular, evaluated at the electroweak scale
\begin{equation}
\beta=\frac{g_{B-L}^{2}}{3} \simeq 0.0358 < 0.14 
\label{acel3}
\end{equation}
suggesting the absense of a bosonic $D_{3}$ condensate in the core of the cosmic string.

\section{Fermionic Superconductivity}

A second possible signature of cosmic strings is superconductivity arising, not from boson condensates but, rather, from zero-modes of charged fermions~\cite{Witten}. 
The relevant fermions are those with bilinear couplings to a scalar field that 1)  has a non-vanishing VEV at radial infinity and 2) which winds non-trivially around the center of the string. 
Exactly which fermions, if any, develop zero-modes is dependent on the theory under consideration and on the explicit cosmic string background~\cite{Weinberg, Witten, PhaseTr}. In the $B$-$L$ MSSM theory described in this paper, the structure and properties of potential zero-modes is very specific.

The cosmic string background described in Section 4 is constructed from the right-handed sneutrino, which has non-vanished VEV at radial infinity and non-zero winding $n$ around the origin. We begin, therefore, by considering the fermions which couple to it.
In the $B$-$L$ MSSM theory, the coupling of the right-handed sneutrino $\nu_{3}$ to charged fermions is completely specified by the superpotential
\begin{equation}
W=\dots + \lambda_{\nu_{3}}L_{3}H\nu_{3} \ ,
\label{night1}
\end{equation}
where $\lambda_{\nu_{3}}$ is the third family neutrino Yukawa coupling of order $10^{-9}$. It follows that the relevant physics is described by
\begin{eqnarray}
&& {\cal{L}}=i{\bar{\psi}}_{LE^{-}_{3}} {\bar{\sigma }}^{\mu} {\cal{D}}_{E \mu} \psi_{LE^{-}_{3}}+i{\bar{\psi}}_{LH^{+}} {\bar{\sigma }}^{\mu} {\partial}_{\mu} \psi_{LH^{+}} \nonumber \\
&& \qquad \qquad - \lambda_{\nu_{3}}( \psi_{LE^{-}_{3}} \psi_{LH^{+}} \nu_{3}+ hc) +\dots \ ,
\label{night2}
\end{eqnarray}
where ${\cal{D}}_{E \mu}=\partial_{\mu}+i g_{B-L}A_{B-L \mu}$. The associated equations of motion are
\begin{eqnarray}
i{\bar{\sigma }}^{\mu} {\cal{D}}_{E \mu} \psi_{LE^{-}_{3}}-   \lambda_{\nu_{3}}  \bar{\psi}_{LH^{+}} \nu^{\ast}_{3} & = &  0  \ , \nonumber \\ 
 i \sigma^{\mu} {\partial}_{\mu} \bar{ \psi}_{LH^{+}}-   \lambda_{\nu_{3}} \psi_{LE^{-}_{3}} \nu_{3} & = & 0 
  \ . 
\label{DiracEq}
\end{eqnarray}
We want to solve these in the background of the cosmic string defined by the transverse functions ${\bf \nu_{3}}$ and ${\bf A_{B-L \mu}}$ in \eqref{buddy5}. Therefore, first consider solutions  of \eqref{DiracEq}
that are independent of time and the $z$-coordinate. Denoting these transverse fermions by $\beta_{LE^{-}_{3}}(x,y) $ and $\beta_{LH^{+}}(x,y)$, equations \eqref{DiracEq} become
\begin{eqnarray}
i{\bar{\sigma}}^{i} {\cal{D}}_{i} \beta_{LE^{-}_{3}} -  \lambda_{\nu_{3}}  \bar{\beta}_{LH^{+}} {\bf\nu^{\ast}_{3}} & = & 0  , \nonumber \\
i \sigma^{i} {\partial}_{i} {\bar{\beta}}_{LH^{+}} - \lambda_{\nu_{3}} \beta_{LE^{-}_{3}} {\bf \nu_{3}} & = & 0 
\label{TransDiracEq1}
\end{eqnarray}
where ${\cal{D}}_{E \mu}=\partial_{\mu}+i g_{B-L}{\bf A_{B-L \mu}}$ and $i=1,2$.
It follows from the index of the corresponding Dirac operator that~\eqref{TransDiracEq1} 
has $|n|$ linearly independent pairs of normalizable zero-modes~\cite{Jackiw, PhaseTr}. These modes are eigenstates of  the $\sigma_{3}$ operator,
\begin{equation}
\sigma^{3}\beta=\frac{n}{|n|} \beta \ .
\label{Chiral}
\end{equation}
Thus, for a given $n$ the zero-modes have the same chirality and each is described by a complex scalar function. Specifically, in cylindrical coordinates the solutions are~\cite{Jackiw, PhaseTr}
\begin{eqnarray}
\beta_{LE^{-}_{3}}(r,\theta) & = & U^{l}_{LE^{-}_{3}} ( r ) e^{i(l-\frac{n}{2}+\frac{1}{2}) \theta} \ , \nonumber\\
\bar{\beta}_{LH^{+}}(r,\theta)& = &U^{l}_{LH^{+}_{3}} ( r ) e^{i(l+\frac{n}{2}-\frac{1}{2}) \theta} 
\label{Chiral2}
\end{eqnarray}
where $-\frac{n}{2}+\frac{1}{2} \le l \le \frac{n}{2}-\frac{1}{2}$. The radial functions can be explicitly evaluated asymptotically. As $r\rightarrow 0$, one finds 
\begin{eqnarray}
 U^{l}_{LE^{-}_{3}} ( r )& \sim & r^{-l+\frac{n}{2}-\frac{1}{2}} \ , \nonumber \\
 U^{l}_{LH^{+}_{3}} ( r )& \sim &r^{l+\frac{n}{2}-\frac{1}{2}}
 \label{radial}
\end{eqnarray}
whereas for $r\rightarrow \infty$ 
\begin{equation}
U^{l}_{LE^{-}_{3}} ( r ), \ U^{l}_{LH^{+}_{3}}(r) \sim e^{-\lambda_{\nu_{3}}<\nu_{3}>r} \ .
\label{radial2}
\end{equation}
Due to the exponential decay, the range of the fermionic solutions is of order 
\begin{equation}
r_{f} \sim \frac{1}{\lambda_{\nu_{3}}\langle\nu_{3}\rangle} \ .
\label{radial3}
\end{equation}
Note from \eqref{buddy9} that for our specific theory the radius of the cosmic string is $r_{s}=r_{v} \sim \frac{1}{g_{B-L}\langle\nu_{3}\rangle}$ and, hence,  
\begin{equation}
\frac{r_{f}}{r_{s}} \sim \frac{g_{B-L}}{\lambda_{\nu_{3}}} \simeq 10^{9} \ .
\label{radial4}
\end{equation}
That is, the radial zero-mode solutions are $10^{9}$ times wider than the vortex core. This indicates the  extremely diffuse nature of the fermionic solutions in the $B$-$L$ MSSM theory.
Any normalizable solution with general boundary conditions is a linear combination of 
these zero-modes. We refer the reader to~\cite{Jackiw, PhaseTr, Weinberg}  for a detailed derivation of these properties. 

Now consider full four-dimensional solutions of~\eqref{DiracEq} of the form
\begin{eqnarray}
\psi_{LE^{-}_{3}}& = & \beta_{LE^{-}_{3}}(x,y) \alpha(z,t) \ ,\\
\bar{\psi}_{LH^{+}}& = & \bar{\beta}_{LH^{+}}(x,y)\alpha(z,t)^{*} \ .
\end{eqnarray}
 Since $\beta_{LE^{-}_{3}}(x,y) $ and $\bar{\beta}_{LH^{+}}(x,y)$ satisfy the transverse Dirac 
 equations~\eqref{TransDiracEq1}, it follows from~\eqref{DiracEq} and~\eqref{Chiral} that
 \begin{equation}
 \Big(\frac{\partial}{\partial t}-\frac{n}{|n|}\frac{\partial}{\partial z}\Big)\alpha(z,t)=0 \ .
\label{radial5}
\end{equation}
Without loss of generality, we henceforth assume that the winding number is positive.
Then \eqref{radial5} implies that $\alpha(z,t)=f(z+t)$.
Thus, an $n>0$ winding of the $\nu_{3}$ scalar in the $B$-$L$ cosmic string 
solution \eqref{buddy5} induces {\it left-moving} fermionic currents in the cosmic string composed of $ \psi_{LE^{-}_{3}} $ and $\psi_{LH^{+}}$.

Anomaly cancellation on the string worldsheet~\cite{Witten} requires that there be chiral fermions which couple to a scalar which winds oppositely to $\nu_{3}$. Since the theory is supersymmetric, this cannot be the conjugate field $\nu_{3}^*$. Furthermore, since the Higgs fields are neutral under $B$-$L$ transformations,  they cannot have topologically stable winding around the string core even though they have non-vanishing VEVs. Note, however, that if the left-handed sneutrino $N_{E_{3}}$ gets an expectation value, then it follows from the equations of motion that this must wind 
oppositely to $\nu_{3}$. Any chiral fermions coupling to the wound solution ${\bf N_{E_{3}}}$ will then generate {\it right-moving} currents on the cosmic string, canceling the anomaly. Does $N_{3}$ have a non-zero expectation value? The answer is affirmative, as we now show.

The neutral scalar fields in the $B$-$L$ MSSM theory are the $H^{0}$, $\bar{H}^{0}$ components of the Higgs fields and the left- and right-handed sneutrinos $N_{i}$, $\nu_{i}$ for $i=1,2,3$. Since $\nu_{3}$ is the only right-handed sneutrino to get a non-zero expectation value, we need only consider the third family. The potential energy of these neutral fields is found to be
\begin{equation}
V_{0}=V_{F}+V_{B-L}+V_{Y}+V_{SU(2)}+V_{soft}
\label{car1}
\end{equation}
where
\begin{eqnarray}
V_{F}=\sum_{m}|F_{m}|^{2} &=& \lambda_{\nu_{3}}^{2}(|\nu_{3}|^{2}|H^{0}|^{2}+|\nu_{3}|^{2}|N_{3}|^{2}+|N_{3}|^{2}|H^{0}|^{2})  \label{car2} \\
&+&\mu^{2}(|H^{0}|^{2}+|\bar{H}^{0}|^{2})
-\lambda_{\nu_{3}}(\mu\nu_{3}N_{3}\bar{H}^{0}+hc) \ , \nonumber 
\end{eqnarray}
\begin{eqnarray}
&& V_{B-L}=\frac{1}{2}D_{B-L}^{2} = \frac{g_{B-L}^{2}}{2}((|\nu_{3}|^{2}-|N_{3}|^{2})^{2}\ , \label{car3} \\ 
&& V_{Y}=\frac{1}{2}D_{Y}^{2} = \frac{g_{Y}^{2}}{2}(|H^{0}|^{2}-|\bar{H}^{0}|^{2}-|N_{3}|^{2})^{2} \ , \label{car4} \\
&& V_{SU(2)}=\frac{1}{2}D_{SU(2)}^{2}=\frac{g_{2}^{2}}{2}(-|H^{0}|^{2}+|\bar{H}^{0}|^{2}+|N_{3}|^{2})^{2} \ , \label{car5}
\end{eqnarray}
\begin{eqnarray}
V_{soft} &=& m_{N_{3}}^{2}|N_{3}|^{2} +m_{\nu_{3}}^{2}|\nu_{3}|^{2}+m_{H}^{2}|H^{0}|^{2}+m_{\bar{H}}^{2}|\bar{H}^{0}|^{2} \label{car6} \\
&-& (BH^{0}{\bar{H}}^{0}+hc)+(A_{\nu_{3}} \nu_{3}N_{3}H^{0}+hc) \ . \nonumber
\end{eqnarray}
Let us solve for the expectation values for each neutral scalar subject to the hierarchy condition
\begin{equation}
\langle N_{3} \rangle \ll \langle H^{0} \rangle , \langle {\bar{H}}^{0} \rangle \ll \langle \nu_{3} \rangle \ .
\label{car7}
\end{equation}
The $\partial V_{0}/\partial \nu_{3}=0$ and $\partial V_{0}/\partial H^{0}=0$, 
$\partial V_{0}/\partial {\bar{H}}^{0}=0$ equations lead to the non-zero expectation values for 
$\langle \nu_{3} \rangle $ and $\langle H^{0} \rangle$ , $\langle {\bar{H}}^{0}\rangle$ presented in \eqref{snow1} and \eqref{snow5} respectively. The $\partial V_{0}/\partial N_{3}=0$ equation then gives
\begin{equation}
\langle N_{3} \rangle = \frac{(\lambda_{\nu_{3}} \mu \langle {\bar{H}}^{0}\rangle-A_{\nu_{3}}\langle H^{0} \rangle)\langle \nu_{3} \rangle}{m_{N_{3}}^{2}-g_{B-L}^{2} \langle \nu_{3} \rangle^{2}} \ .
\label{car8}
\end{equation}
Therefore, following the electroweak phase transition the left-handed sneutrino acquires a very small expectation value. For example, assuming $\mu \sim \langle H^{0} \rangle$, $A_{\nu_{3}} \sim \lambda_{\nu_{3}} \langle H^{0} \rangle$ and $m_{N_{3}}^{2}-g_{B-L}^{2} \langle \nu_{3} \rangle^{2} \sim g_{B-L}^{2} \langle \nu_{3} \rangle^{2}$, it follows that
\begin{equation}
\langle N_{3} \rangle \sim (10^{-10}-10^{-12}) \langle \nu_{3} \rangle
\label{fri1}
\end{equation}
for the $B$-$L$/electroweak hierarchy given in \eqref{snow10a}. That is, $\langle N_{3} \rangle$ is on the order of the neutrino masses.
This is sufficient, however, to provide the right-moving fermionic zero-modes on the cosmic string required by anomaly cancellation. Note that the vanishing of $\langle \nu_{3} \rangle$ at the center of the cosmic string will set $\langle N_{3} \rangle=0$, consistent with a non-trivial winding of ${\bf N_{3}}$.

To see how these arise, note that the coupling of the left-handed sneutrino $N_{3}$ to charged fermions is specified by the superpotential
\begin{equation}
W=\dots + \lambda_{e_{3}}L_{3}\bar{H} e_{3} \ ,
\label{night1}
\end{equation}
where $\lambda_{e_{3}}$ is the third family $\tau$ Yukawa coupling of order $5\times10^{-2}$. The relevant physics is then described by
\begin{eqnarray}
&& {\cal{L}}=i{\bar{\psi}}_{Le^{+}_{3}} {\bar{\sigma }}^{\mu} {\cal{D}}_{e \mu} \psi_{Le^{+}_{3}}+i{\bar{\psi}}_{LH^{-}} {\bar{\sigma }}^{\mu} {\partial}_{\mu} \psi_{LH^{-}} \nonumber \\
&& \qquad \qquad - \lambda_{e_{3}}( \psi_{Le^{+}_{3}} \psi_{LH^{-}} N_{3}+ hc) +\dots \ ,
\label{car9}
\end{eqnarray}
where ${\cal{D}}_{e \mu}=\partial_{\mu}-i g_{B-L}A_{B-L \mu}$. It follows from the above analysis that with respect to the cosmic string defined by ${\bf \nu_{3}}$, ${\bf A_{B-L \mu}}$ in \eqref{buddy5} and the associated background ${\bf N_{3}}$, there will be $|n|$ linearly independent pairs  $\beta_{Le^{+}_{3}}(x,y)$, $\beta_{LH^{-}}(x,y)$
of transverse normalizable fermion zero-modes. These modes are eigenstates of $\sigma_{3}$ and
have a structure similar to \eqref{Chiral2}. However, whereas the solutions in \eqref{Chiral2} and their $\sigma_{3}$ eigenvalue are indexed by $n$, the winding of ${\bf \nu_{3} }$, these zero-modes are indexed by $-n$, the winding of ${\bf N_{3}}$. The small $r$ behaviour remains similar to that in \eqref{radial}. Now, however, as $r \rightarrow \infty$
\begin{equation}
U^{l}_{Le^{+}_{3}} ( r ), \ U^{l}_{LH^{-}_{3}}(r) \sim e^{-\lambda_{e_{3}}<N_{3}>r} 
\label{car10}
\end{equation}
and, hence, the range of these fermionic solutions is of order 
\begin{equation}
r_{F} \sim \frac{1}{\lambda_{e_{3}}\langle N_{3}\rangle} \ .
\label{car11}
\end{equation}
It follows from this, \eqref{radial3} and \eqref{fri1} that
\begin{equation}
r_{F} \sim (10^{3}-10^{5}) r_{f} \ .
\label{fri2}
\end{equation}
Therefore, these radial zero-mode solutions are even more diffuse around the cosmic string core.
The full four-dimensional solutions are again of the form $\psi_{L}=\beta_{L}(x,y)\alpha(z,t)$. Now, however, the function $\alpha$ satisfies
\begin{equation}
\Big(\frac{\partial}{\partial t}+\frac{\partial}{\partial z}\Big)\alpha(z,t)=0 \ ,
\label{car12}
\end{equation}
implying that  $\alpha(z,t)=f(z-t)$.
Thus, the $n<0$ winding of the ${\bf N_{3}}$ scalar in the background of  \eqref{buddy5} induces {\it right-moving} fermionic currents in the cosmic string composed of 
$ \psi_{Le^{+}_{3}} $ and $\psi_{LH^{-}}$.

Having found both the left- and right-moving charged fermionic modes of a $B$-$L$ MSSM cosmic string, we want to analyze whether they can lead to cosmologically observable phenomenon and, specifically, to superconductivity.
As can be seen from the above discussion, the electroweak phase transition is necessary to 
produce the right-moving modes to cancel the anomaly. However, the superpotential term
in \eqref{night1} will then generate, in addition to the fermion coupling to $N_{3}$ discussed above, a Yukawa mass $\lambda_{e_{3}} \psi_{LE_{3}^{-}} \langle {\bar{H}}^{0} \rangle \psi_{Le_{3}^{+}}$ for the tauon. In addition, 
 a non-zero $\mu$-term $\mu H \bar{H}$ is required in the superpotential to make Higgsinos massive. Specifically, a mass term of the form $\mu \psi_{H^{+}} \psi_{H^{-}}$ will appear. Both masses are much larger than neutrino masses and, as a result, the zero-modes
of the previous discussion will be lifted; generically, turning into massive bound states~\cite{HillWidow,  Hindmarsh, PhaseTr, DavisR} with mass $m_{\tau}=1.776~GeV$ and 
$m_{Higgsino}\sim \mu$ respectively. For an applied electric field in the string frame satisfying $E\ll 2 \pi m^{2}$, massive bound states cannot form persistent currents. Their maximum electric current is given by~\cite{HillWidow}
\begin{equation}
J_{max}\approx \frac{E}{2\pi^{3/2}m} \ .
\label{feather1}
\end{equation}
Note that $J_{max}$
is directly proportional to the applied field. As soon as $E=0$, this charged fermionic current will relax to zero. The maximal such currents generated in a $B$-$L$ MSSM cosmic string by tauon bound states, for example, would be of order $10^{6}$-$10^{7}A$. For  $E \gg 2 \pi m^{2}$, it is possible to obtain currents in this setting which are close to superconducting. However, for that to happen one needs to be in the regime 
\begin{equation}
B ( v/c ) \ge 10^{3}\big(m/1 \text{eV}\big)^{2}~Gauss \ , 
\label{feather2}
\end{equation}
where $v$ is the transverse string velocity. Hence, for a superconducting current of tauons, the required magnetic field would be at least of order $10^{21}~ Gauss$, far larger than any observed cosmological $B$ field. Thus, fermionic currents of the $B$-$L$ MSSM cosmic string are unlikely to have any observable cosmological effects.

\section*{Acknowledgements} We would like to thank J. Khoury, N. Turok, T. Vachaspati, A. Vilenkin, M. Trodden and D. Wesley for helpful discussion.
T.~Brelidze and B.~A.~Ovrut are supported in part by the DOE under contract No.
DE-AC02-76-ER-03071 and by NSF RTG Grant DMS-0636606.


\section*{Appendix: Numerical Analysis of the Stability Equation}

In this Appendix, we present a numerical procedure for determining the existence of negative eigenvalue, normalizable solutions of the stability equation
\begin{equation}
\label{Schroed1}
( -\frac{\partial^{2}}{\partial r^{2}} -\frac{1}{r}\frac{\partial}{\partial r}){\bf \sigma}_{0}+ {\hat{V}}{\bf \sigma}_{0}=\omega^{2}{\bf \sigma}_{0} \ ,
\end{equation}
where
\begin{equation}
{\hat{V}}(r)=\beta \eta_{\phi}^{2}f(r)^{2}-\frac{\lambda_{\sigma}\eta_{\sigma}^{2}}{2}+
n^{2}(\frac{\tilde{q}_{\sigma}^{2}}{\tilde{q}_{\phi}^{2}}) \frac{\alpha(r)^{2}}{r^{2}} \ .
\label{Schroed2}
\end{equation}
Although applicable in general, we specify our algorithm for the simplest case discussed in the text where $n=1$, $\tilde{q}_{\sigma}^{2}=\tilde{q}_{\phi}^{2}$ and $m_{\sigma}^{2}=0$, resulting in 
\eqref{ny2}. Potential \eqref{Schroed2} then simplifies to
\begin{equation}
{\hat{V}}(r)=\beta (f(r)^{2}-1)+\frac{\alpha(r)^{2}}{r^{2}} \ ,
\label{Schroed3} 
\end{equation}
where we have set $\eta_{\phi}=1$ and, hence, the radial coordinate $r$ and parameter $\beta$ are dimensionless. Note that ${\hat{V}}$ depends only on the single parameter $\beta$.
Furthermore, impose the critical coupling constraint $\frac{\lambda_{\phi}}{2\tilde{g}^{2}}=1$,
thus simplifying the functions $f(r)$ and $\alpha(r)$ to be solutions of \eqref{buddy11}. These equations have been solved numerically in the literature~\cite{Vilenkin} and we input their solutions directly into our analysis of the stability equation.

To prove the existence of a boson condensate for a fixed value of parameter $\beta$ in \eqref{Schroed3}, it suffices to find a negative energy ground state solution to \eqref{Schroed1}. Hence, one can impose the boundary
conditions 
\begin{equation}
\label{initial}
\sigma_{0}\Big|_{r=0}= 1 \ ,  \quad \qquad \partial_{r} \sigma_{0} \Big|_{r=0} = 0 \ .
\end{equation}
In addition, to ensure that $\sigma_{0}(r)$ is normalizable constrain
\begin{equation}
\sigma_{0}\Big|_{r \rightarrow \infty}=0 \ .
\label{norm}
\end{equation}
Note that if a bound state  exists, its eigenvalue can never be more negative than the depth of the potential energy. Hence,
the possible range of values for $\omega^{2}$ is limited to 
\begin{equation}
-\beta<\omega^{2}<0.
\label{snow1}
\end{equation}
For each such $\omega^{2}$, there is a solution $\sigma_{0}$ which satisfies~\eqref{Schroed1}
with boundary conditions~\eqref{initial}. Generically, however, this solution will not be normalizable. 
To see this, note that for large $r$ the solution  to \eqref{Schroed1} must be of the form
 \begin{equation}
 \sigma_{0}(r)\stackrel{r \rightarrow \infty}{\longrightarrow} C_{1}r^{-1/2}e^{-|\omega| r}+ C_{2}r^{-1/2}e^{|\omega| r},
 \end{equation}
where $C_{1}$ and $C_{2}$ are continuous functions of $\omega^{2}$. If for a chosen value of $\omega^{2}$ $C_{2}$ is non-vanishing, then $\sigma_{0}$ diverges at large $r$ and the renormabiliability constraint \eqref{norm} is not satisfied. Note that $C_{2}\neq0$ can be either positive of negative. If positive, $\sigma_{0} \stackrel{r \rightarrow \infty}{\longrightarrow}+\infty$, that is, the wavefunction ``flips up'' at large $r$. On the other hand, if $C_{2}$ is negative,  $\sigma_{0} \stackrel{r \rightarrow \infty}{\longrightarrow}-\infty$ and the wavefunction ``flips down''. It is only if $C_{2}$ exactly vanishes for a specific value $\omega_{0}^{2}$ in \eqref{snow1}, that constraint \eqref{norm} is satisfied and the wavefunction normalizable.

\begin{figure}[p]
\centering
\includegraphics[scale=0.5]{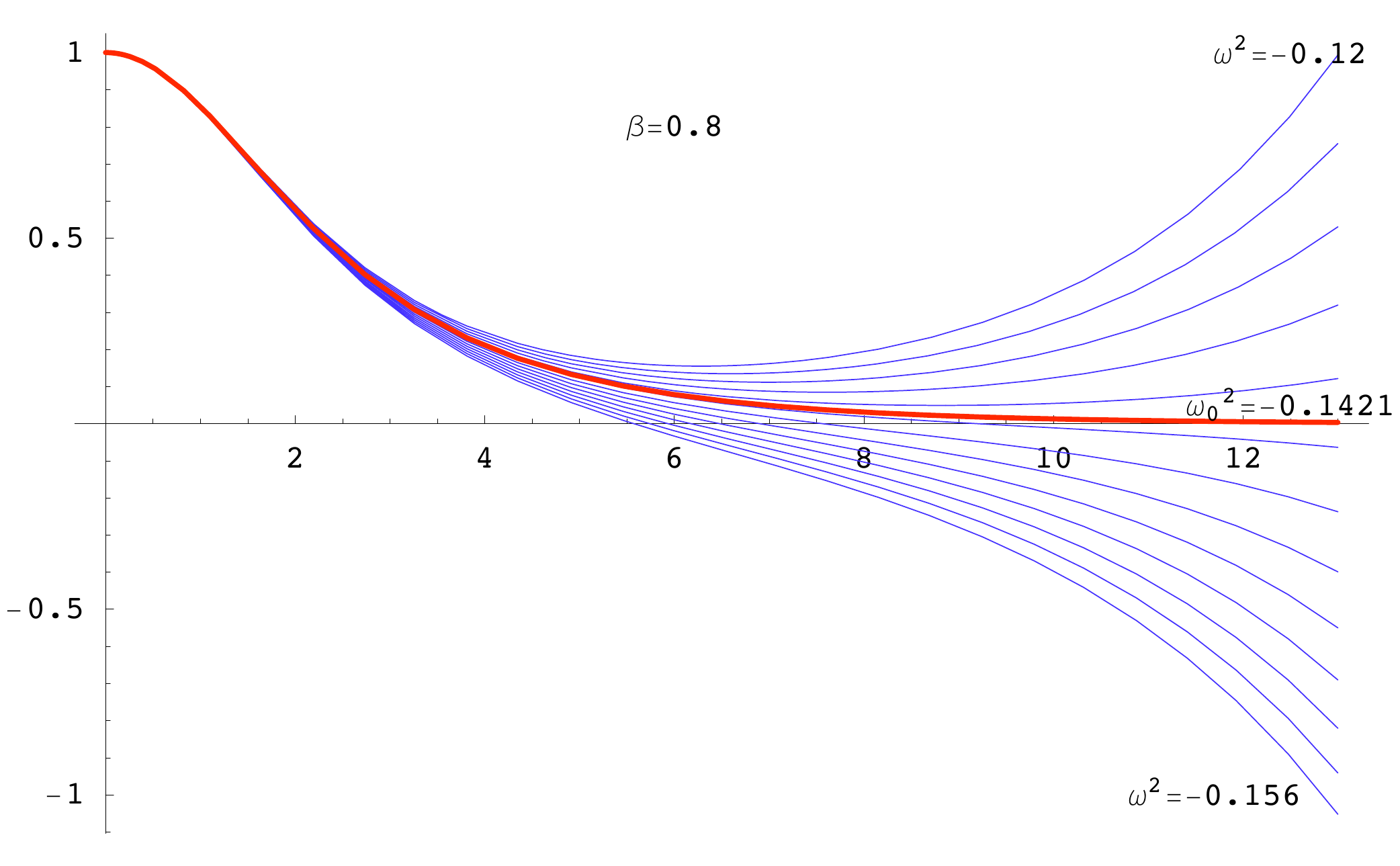}
\caption{Family of $\sigma_{0}$ solutions for the initial value problem with $\beta=0.8$ and 
$\omega^{2}$ varying from -0.12
to -0.156. Note that the asymptotic behaviour of the wavefunction changes sign. 
The ground state occurs at $\omega_{0}^{2}=-0.1421$ and its associated normalizable 
ground state is indicated in red. }
\label{Fig2}
\end{figure}

Two scenarios are then possible. First, if when $\omega^{2}$ is varied over the entire range \eqref{snow1} $C_{2}$ is always greater than, or always less than,  zero, then the wavefunction is never normalizable and a negative eigenvalue ground state does not exist for this choice of parameter $\beta$. Second, if when $\omega^{2}$ is varied over range \eqref{snow1} $C_{2}$ changes sign, then there must be an $\omega_{0}^{2}$ for which
\begin{equation}
C(\omega_{0}^{2})=0 \ ,
\label{snow2}
\end{equation}
since $C_{2}$ is a continuous function of $\omega^{2}$. Hence, for this choice of $\beta$ a normalizable ground state solution for $\sigma_{0}$ exists with negative energy $\omega_{0}^{2}$. These results give us an explicit algorithm for computing the existence, or non-existence, of a boson condensate. This is:

\begin{enumerate}

\item Choose a fixed value for parameter $\beta$.

\item Vary $\omega^{2}$ over the range \eqref{snow1}. 

\item For each value of $\omega^{2}$, numerically solve \eqref{Schroed1},\eqref{Schroed3} for the ground state wavefunction $\sigma_{0}$ satisfying boundary conditions \eqref{initial}. We do this by  implementing the Runge-Kutta method on Mathematica. 

\item Plot $\sigma_{0}$ versus $r$  for all values of $\omega^{2}$. 

\item If all these curves ``flip up'' or ``flip down'', then there is no negative energy ground state. However, if these curves ``flip up'' for small values of $\omega^{2}$ but ``flip down'' for larger values, then a ground state with negative energy $\omega_{0}^{2}$ does exist. 

\item To compute $\omega_{0}^{2}$ and the associated normalizable wavefunction, we numerically identify
the interval which contains $\omega^{2}_{0}$. We then iterate this procedure until we obtain the ground
state energy and wavefunction to the desired precision, thus approximating the solution to the stability equation. 

\end{enumerate}
To make this concrete, in~\autoref{Fig2} we carry out this algorithm explicitly for parameter $\beta=0.8$.
Observe that $\sigma_{0}$ ``flips up'' for small $\omega^{2}$, but ``flips down'' for larger values of 
$\omega^{2}$. This signals the existence of a negative energy ground state occurring in between, when $\sigma_{0} \stackrel{r \rightarrow \infty}{\longrightarrow} 0$. The numerical value of $\omega_{0}^{2}=-0.1421$ and the normalizable wavefunction are both indicated in the Figure. Note that 
\begin{equation}
\frac{|\omega_{0}^{2}|}{\beta}= 0.1776 \ ,
\label{snow3}
\end{equation}
that is, the bound state energy is $17.76$ \% of the depth of the potential.

\begin{figure}[p]
\centering
\includegraphics[scale=0.5]{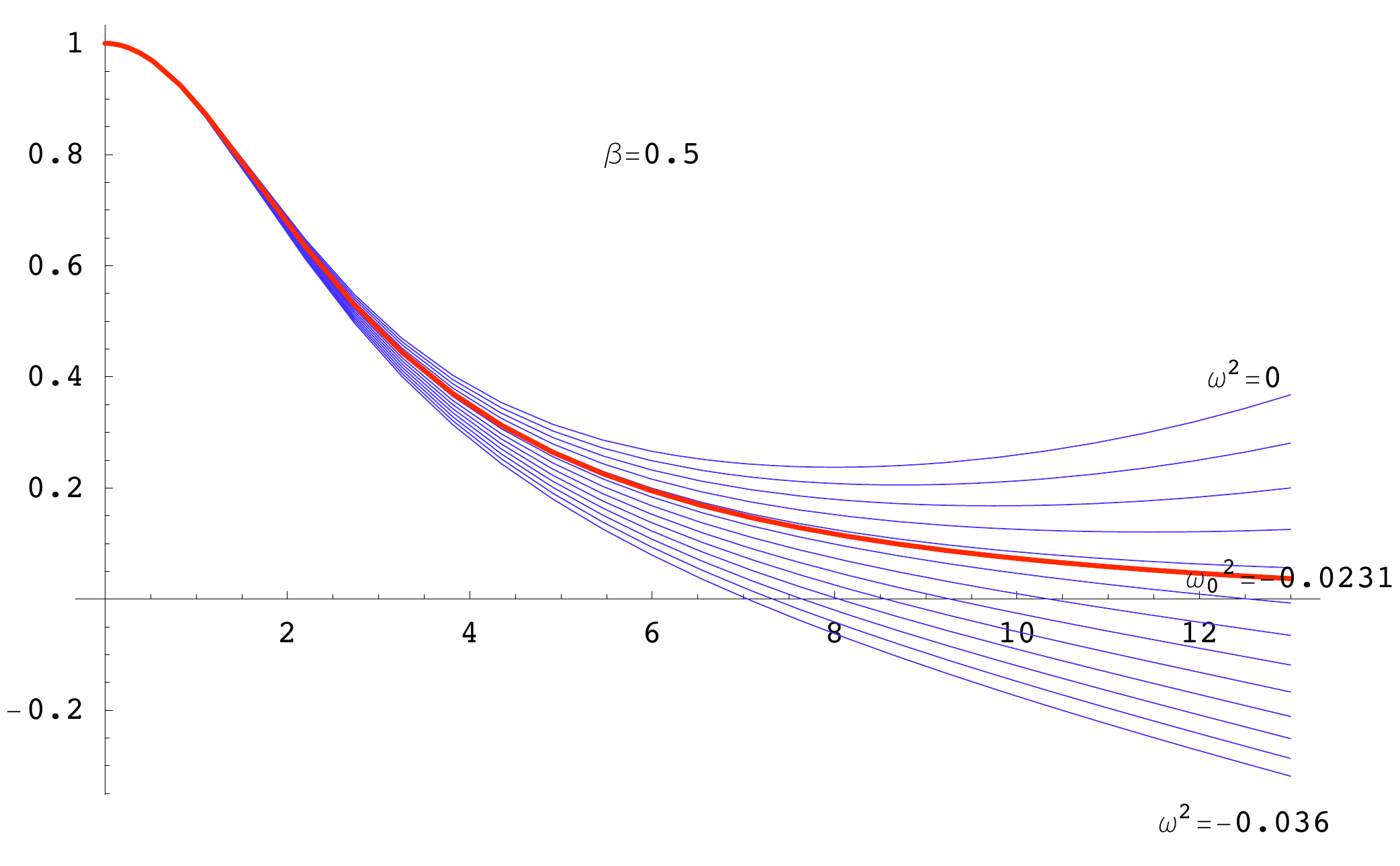}
\caption{Family of $\sigma_{0}$ solutions for the initial value problem with $\beta=0.5$ and $\omega^{2}$ varying from -0.003
to -0.036. Note the changing sign in the asymptotic behaviour of the wavefunction. The ground state occurs at $\omega_{0}^{2}=-0.0231$ and the associated normalizable ground state is indicated in red.}
\label{Fig3}
\end{figure}

\begin{figure}[p]
\centering
\includegraphics[scale=0.5]{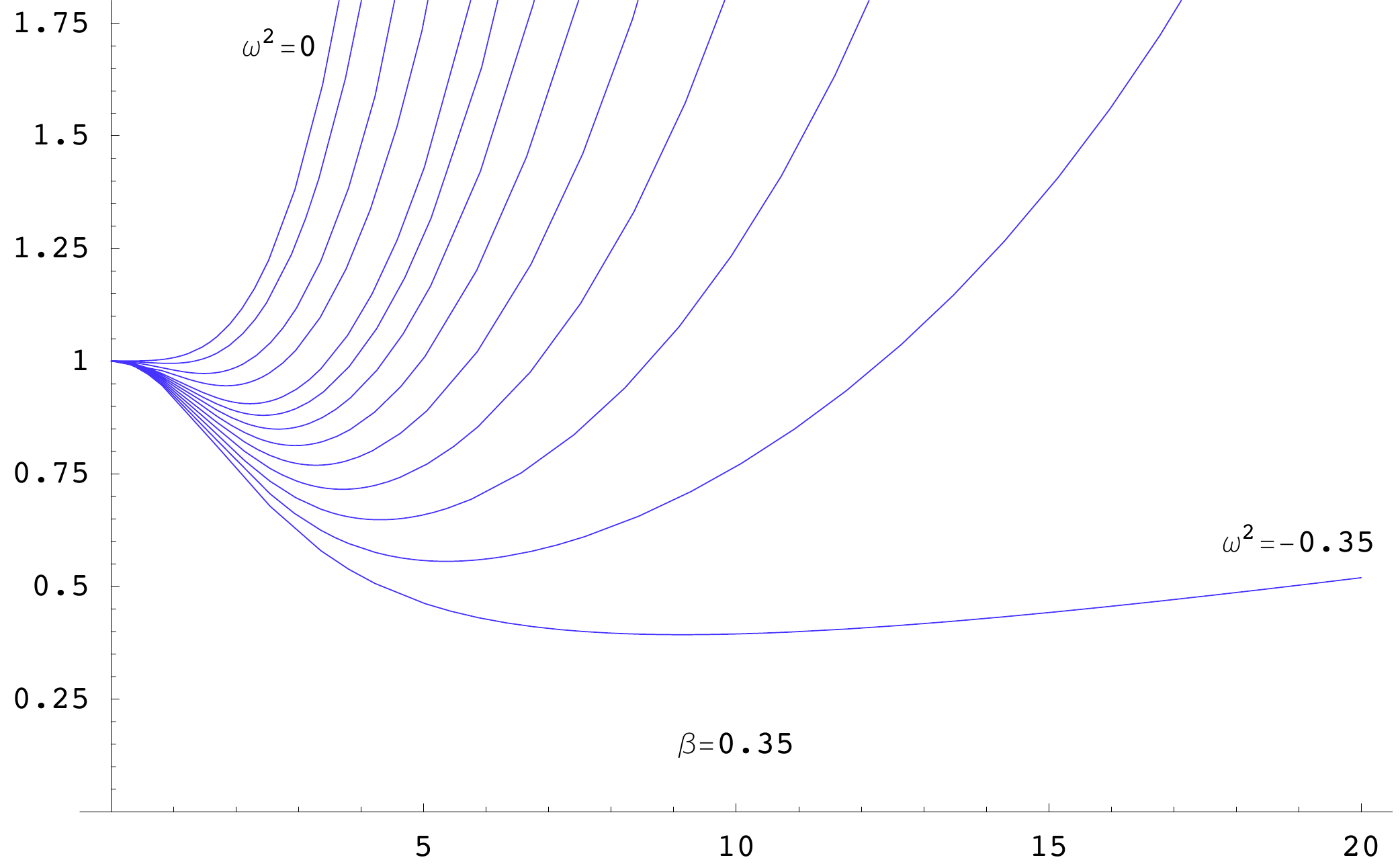}
\caption{Family of $\sigma_{0}$ solutions for the initial value problem with $\beta=0.35$ over the entire allowed range of $\omega^{2}$. Note that the asymptotic values of the wavefunctions are always positive, diverging to $+\infty$. This corresponds to the stability equation admitting no negative energy ground state.}
\label{Fig4}
\end{figure}

\begin{figure}[p]
\centering
\includegraphics[scale=0.5]{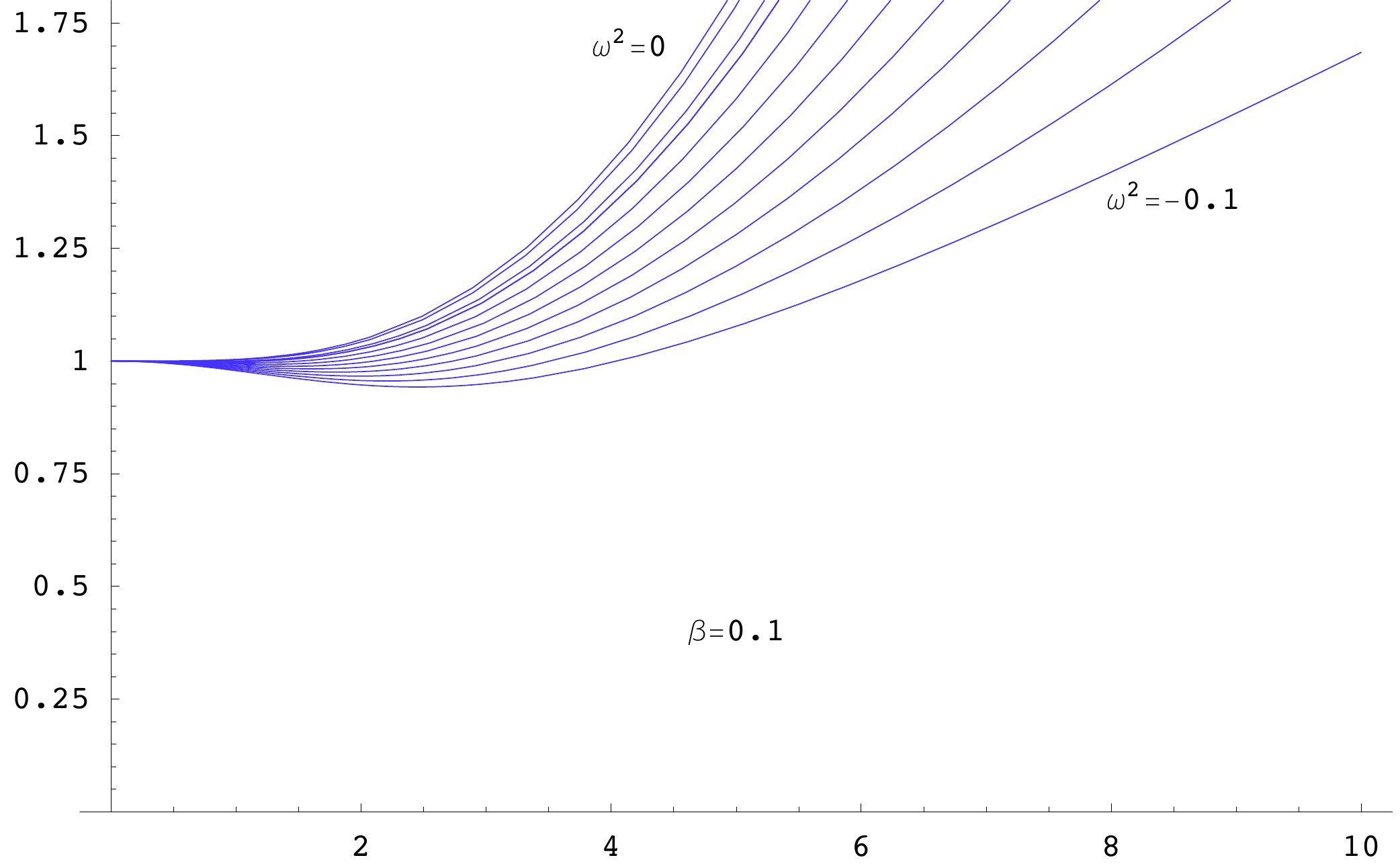}
\caption{Family of $\sigma_{0}$ solutions for the initial value problem with $\beta=0.1$ over the entire allowed range of $\omega^{2}$. Note that the asymptotic values of the wavefunctions are always positive, diverging to $+\infty$. This corresponds to the stability equation admitting no negative energy ground state.}
\label{Fig5}
\end{figure}

Let us now carry out this computation for smaller values of $\beta$. 
The results for $\beta=0.5$ are shown in~\autoref{Fig3}. 
Again, note that $\sigma_{0}$ ``flips up'' for small $\omega^{2}$, but ``flips down'' for larger values of 
$\omega^{2}$. This signals the existence of a negative energy ground state occurring in between, when $\sigma_{0} \stackrel{r \rightarrow \infty}{\longrightarrow} 0$. The numerical value of $\omega_{0}^{2}=-0.0231$ and the normalizable wavefunction are both indicated in the Figure. In this case,
\begin{equation}
\frac{|\omega_{0}^{2}|}{\beta}= 0.0462 \ ,
\label{snow3}
\end{equation}
that is, the bound state energy is $4.62$ \% of the depth of the potential.
Note that the percentage size of the eigenvalue relative to the depth of the potential has substantially decreased over the $\beta=0.8$ case above. This indicates that for some value of $\beta$ not too much smaller than $0.5$ a negative energy ground state might cease to exist. To explore this further, we apply our algorithm to a range of values of parameter $\beta$. 
The ground state energy for each $\beta$,
as well as their fractional depth with respect to the potential,  are shown in~\autoref{table}. 
Note that as $\beta$ approaches $\sim 0.42$, $\omega_{0}^{2} \rightarrow 0$ and is a rapidly decreasing percentage of the potential depth. 
Indeed, we find that for
\begin{equation}
\beta<\beta_{critical} \simeq 0.42 \ ,
\label{critical}
\end{equation}
there is no negative energy bound state solution to \eqref{Schroed1},\eqref{Schroed3}. Two concrete examples of this are $\beta=0.35$ and $\beta=0.1$. Our numerical results for these parameters are shown in~\autoref{Fig4} and~\autoref{Fig5} respectively. For both cases we see that, unlike the previous examples, $\sigma_{0}$ always ``flips up '' for all values of $\omega^{2}$ satisfying 
\eqref{snow1}. It follows that in each case there is no negative energy ground state solution to the stability equation. To conclude: we have shown numerically that the stability equation 
\eqref{Schroed1} with potential \eqref{Schroed3} admits a negative energy ground state normalizable solution if and only if
\begin{equation}
\beta> \beta_{critical} \simeq 0.42 \ .
\label{final}
\end{equation}

 \renewcommand{\arraystretch}{1.65}
\begin{table}[]
\begin{center}
\begin{tabular}{|l|c|l|} 
\hline
$\beta$ & $|\omega_{0}|^{2}$&$ |\omega_{0}|^2/ \beta$ \\ \hline $1$ & $0.2404$ & $0.2404$ \\ $0.8$ &$ 0.1421$ & $0.1776$ \\ $0.7$ & $0.0977$ & $0.1396$ \\ $0.5$ & $0.0231$ & $0.0462$ \\ $0.45$ & $0.0094$ & $0.0209$ \\ $0.42$ & $0.0027$ & $0.0064$\\ 
 \hline
 \end{tabular} 
\end{center}
\caption{The ground state energy corresponding to different values of $\beta$. Note that as 
the potential becomes more shallow, the ground state energy decreases relative to the depth of the potential.}
\label{table}
\end{table}

 As discussed in the text, a similar analysis must be carried out with the charges chosen to be $\tilde{q}_{\phi}^{2}=9\tilde{q}_{\sigma}^{2}$. This changes potential \eqref{Schroed3} to
\begin{equation}
{\hat{V}}(r)=\beta(f(r)^{2}-1)+\frac{\alpha(r)^{2}}{9r^{2}} \ .
\label{secpot}
\end{equation}
The numerical analysis of this case gives the same qualitative results, so we won't present it here. Suffice it to say that, due to the weaker repulsion term 
in the potential, the critical value for $\beta$ is lowered. Specifically, we find that the stability equation \eqref{Schroed1} with potential \eqref{secpot} will admit a negative energy normalizable ground state if and only if 
\begin{equation}
\beta > \beta_{critical}\simeq 0.14 \ .
\label{final1}
\end{equation}
%


\end{document}